\documentclass[preprint,showpacs,preprintnumbers,amsmath,amssymb,showkeys]{revtex4}

\usepackage{epsfig}
\usepackage{dcolumn}
\usepackage{bm}

  \newcommand{\nn}{\nonumber}
  \newcommand{\sura}{\hspace{-1.5mm}\!/}
  
\begin{document}


\title{Jet substructures of boosted polarized hadronic tops}

\author{Yoshio Kitadono}
\email{kitadono@phys.sinica.edu.tw}

\author{Hsiang-nan Li}
 \email{hnli@phys.sinica.edu.tw}
 \affiliation{Institute of Physics, Academia Sinica, Taipei 11529, Taiwan, Republic of China}

\date{\today}

\begin{abstract}
We study jet substructures of a boosted polarized top quark, which
undergoes the hadronic decay $t\to b u\bar d$, in the perturbative QCD
framework, focusing on the energy profile and the
differential energy profile. These substructures are factorized into the
convolution of a hard top-quark decay kernel with a bottom-quark
jet function and a $W$-boson jet function, where the latter is
further factorized into the convolution of a hard $W$-boson decay kernel
with two light-quark jet functions. Computing the hard kernels to leading order
in QCD and including the resummation effect in the jet functions,
we show that the differential jet energy profile is a useful observable
for differentiating the helicity of a boosted hadronic top quark: a right-handed top
jet exhibits quick descent of the differential energy profile with the inner test
cone radius $r$, which is attributed to the $\mbox{V-A}$ structure of weak
interaction and the dead-cone effect associated with the $W$-boson jet.
The above helicity differentiation may help to reveal the chiral structure of physics
beyond the Standard Model at high energies.

\end{abstract}



\maketitle

\section{Introduction}

Precise theoretical and experimental investigations of
top-quark properties \cite{top.review1, top.review2,top.review3, top.review4} are
crucial not only for understanding the electroweak dynamics in
the Standard Model, but also for exploring new physics beyond the Standard
Model. A top quark may be produced with large boost
at the Large Hadron Collider (LHC) in the future 13-14 TeV run, for example,
directly through the gluon-gluon fusion, or indirectly through
decays of new massive particles from Kaluza-Klein gluons \cite{boost.top.motiv1, boost.top.motiv2},
string Regge states in the Randall-Sundrum model \cite{boost.top.motiv3} and
the supersymmetric models \cite{boost.top.motiv4, boost.top.motiv5}, or light quark partners
in the composite Higgs models \cite{boost.top.motiv6, boost.top.motiv7, boost.top.motiv8,
boost.top.motiv9, boost.top.motiv10, boost.top.motiv11}.
Fixed-order calculations and soft-gluon resummations associated with the boosted top
quark production have been performed in the Standard Model
\cite{boost.top.SM1, boost.top.SM2,boost.top.SM3,boost.top.SM4}.
New physics beyond the Standard Model has been
explored intensively in \cite{boost.top.BSM1, boost.top.BSM2,boost.top.BSM3,
boost.top.BSM4, boost.top.BSM5, boost.top.BSM6, boost.top.BSM7,
boost.top.BSM8} and model independently in \cite{general.Wtb.boosted.top}
by means of boosted top quarks.

It is known that chirality becomes equivalent to helicity for a highly boosted
particle, so the helicity information of an energetic top quark may reveal the chiral
structure of new physics, such as the chiral couplings of a top quark to new
physics. However, the decay products of a boosted top quark may be extremely collimated to each other
and form a single jet. How to differentiate a boosted top quark from ordinary QCD jets,
and then to determine its helicity become a challenge at high energy colliders.
Various strategies for tagging
boosted top quarks were proposed in \cite{boost.top.tagg1, boost.top.tagg2, boost.top.tagg3,
boost.top.tagg4, boost.top.tagg5, boost.top.tagg6, boost.top.tagg7, boost.top.tagg8},
and studied recently in \cite{hyper.boosted.top.tagger, dipole.ultra.boosted.top,
ANN.top.tagger, better.boosted.top.tagger}.
Relevant experimental investigations have been conducted thoroughly by the ATLAS and
CMS Collaborations in \cite{boost.top.exp1, boost.top.exp2}.

It has been shown that the distribution
in the top jet invariant mass is not sensitive to the helicity, since
its peak position is basically determined by the top-quark mass \cite{Kitadono:2014hna}.
Additional information on the internal structure of a top jet
is then needed to distinguish its helicity. The relation between the helicity of
a boosted top quark and the energy fraction distribution of a particular subjet
was discussed in \cite{boost.top.pol} using
Monte Carlo generators. Given an algorithm for the subjet selection,
different energy fraction distributions for the left- and right-handed
top jets were derived. The technique of measuring the top-quark polarization
in the hadronic channel was improved with the weight method in \cite{boost.top.pol2}.
The above progress implies that jet substructures can serve as efficient
observables for distinguishing the helicity of a boosted polarized top quark.
Following this development, we have demonstrated in the perturbative QCD (pQCD)
framework \cite{Kitadono:2014hna} that the energy profile (jet shape)
\cite{energyprofile.definition} of a semi-leptonic top jet
is a simple and useful substructure for the helicity identification, which
does not require decomposition of subjets and algorithms for subject
selection as in \cite{boost.top.pol}, $b$-tagging, $W$-reconstruction or
measurement of missing momenta.

The analysis in \cite{Kitadono:2014hna} began with the construction of a jet
energy function for a polarized top quark, which undergoes the semi-leptonic
decay $t\to b\ell\nu$. The lepton energy is not included, and the neutrino
energy, as a missing momentum, does not contribute either. This jet energy
function was factorized into the convolution of a hard top-quark decay kernel
with a bottom jet energy function. The latter is well approximated by the
light-quark jet energy function derived in the QCD resummation formalism
\cite{energyprofile}, as the jet momentum is high enough. Evaluating the hard
kernel to leading order (LO) in QCD, we obtained the dependencies of the left-
and right-handed top-quark jet energy functions on the top-jet momentum, the
jet cone radius $R$, and the radius $r\le R$ of a test cone centered around the
top jet axis. Normalizing the jet energy functions to their values at $r=R$, we
predicted the energy profiles $\Psi(r)$ of the left- and right-handed top jets.
It was found that the energy profile is sensitive to the helicity for the top jet
momentum around 1 TeV: energy is accumulated faster within a
left-handed top jet than within a right-handed one.

In this paper we will extend our formalism to the study of the energy profiles
of a boosted hadronic top quark, which undergoes the $t\to bu\bar d$ decay, and
explore the dependence of this jet substructure on the top-quark helicity. Compared
to the semi-leptonic top jet, several theoretical challenges have to be overcome.
The kinematics for the three-body decay $t \to bu\bar d$ is much more complicated,
so that it is not easy to handle the angular relation among the three subjets formed
by the bottom, up, and down quarks, as their energy contributions to the test cone
are evaluated. With the neutrino momentum being integrated out, the semi-leptonic
decay involves basically two-body kinematics. For a hadronic top jet, an additional
nonperturbtive function must be introduced to absorb soft gluon exchanges among
the three subjets. Even when we consider a fat bottom jet as in the semi-leptonic case
\cite{Kitadono:2014hna}, which absorbs soft gluons emitted by the top and bottom
quarks, soft gluon exchanges remain between, for example, the up and down quarks.
At last, when subjets overlap largely, their factorization becomes questionable,
and the estimate of their contributions to the test cone becomes tedious in order to avoid
double counting. Such a subjet merging issue does not exist in the semi-leptonic case,
because only a single bottom jet appears in the final state. Another ambiguity related to
the subjet merging arises from the choices of the subjet cone radii, which complicates
the organization of large logarithms involved in the jet energy profiles.

The aforementioned theoretical challenges can be overcome by the sequential factorization
procedures proposed in the present work. We start with a polarized hadronic top jet, and
factorize it into the convolution of a hard top-quark decay kernel with a fat $W$-boson jet
and a fat bottom jet. The fat $W$-boson jet is further factorized into the convolution
of a $W$-boson decay kernel with a fat light-quark jet and a thin light-quark jet.
At each step of factorization, we just need to handle two-body kinematics of final states.
The fat bottom jet absorbs the soft gluons the same as in the semi-leptonic case, namely,
the bottom jet is universal for the semi-leptonic and hadronic top jets.
The fat light-quark jet in the $W$-boson jet absorbs the rest of
soft gluons emitted within the top jet, following the procedure in
\cite{boost.higgs.subjet} that is applied to the absorption of soft gluons
in a Higgs-boson jet. Soft gluon exchanges between the color-singlet $W$-boson jet
and other subprocesses are expected to be suppressed in the limit of high
jet energy. That is, soft gluons are handled by constructing fat subjets, instead
of by introducing additional nonperturbative functions. The fat light-quark jet and
the thin light-quark jet completely overlap, such that the subjet merging issue does not
exist. Similarly, the fat $W$-boson jet and the fat bottom jet completely overlap as well.
In our construction a fat subjet has radius $R$ to absorb all soft radiations in the top jet,
and a thin subjet has radius $r$, which is regarded as a small parameter in our formalism.
We focus on the behavior of the energy profiles
at small $r$, where the resummation technique applies. The ambiguity
to define jet radii is then removed.

Including the resummation effect \cite{energyprofile} in the quark jet functions, 
and evaluating the LO hard kernel under the narrow-width
approximation for both the top-quark and $W$-boson propagators, we derive
the left-handed (helicity-minus) and right-handed (helicity-plus)
top jet energy functions.
It is observed that the bottom jet contributes more to the energy profile of
a left-handed top jet, similar to the semi-leptonic case. As explained in
\cite{Kitadono:2014hna}, this is a consequence attributed to the $\mbox{V-A}$ structure of
weak interaction. The contribution of the $W$-boson jet, exhibiting a more obvious
dead-cone effect, starts from a larger test cone radius, and is more dominant in
the energy profile of a right-handed top jet. Combining the bottom and $W$-boson
jets, we find that the energy profile becomes insensitive to the top-quark
helicity. Instead, the differential energy profile shows a more significant
dependence on the top-quark helicity: the differential energy profile of a
right-handed top jet descends more quickly with the test cone radius $r$.
Our work does not only represent an extension of pQCD to the study of jet
substructures of a boosted weakly decaying massive particle, but also manifests the
differential energy profile as a simple and useful
observable for distinguishing the helicity of a boosted hadronic top quark.

The pQCD factorizations for the jet function and the jet energy function of
a polarized hadronic top quark are formulated in Sec.~\ref{formalism}.
The sensitivity of the energy profile and the differential energy profile of a
hadronic top jet to the helicity is explored numerically in Sec.~\ref{result}.
Section~\ref{conclusion} contains the conclusion. In the Appendix we explain
at one-loop level that soft gluon emissions in a hadronic top jet can be absorbed
into fat subjets.

\section{Formalism \label{formalism}}

In this section we derive the factorization formulas for the jet function
and the jet energy function
of a highly boosted polarized hadronic top quark. The jet energy profile
\begin{eqnarray}
\Psi(r) &=&
\frac{1}{N_{J_t}}\sum_{J_t}\frac{\sum_{\theta_i<r, i\in J_t}P_{T_i}}
{\sum_{\theta_i<R, i\in J_t}P_{T_i}},\label{profile}
\end{eqnarray}
is then predicted, in which $N_{J_t}$ is the number of top jets with the cone radius $R$,
$r$ is the test cone radius, $P_{T_i}$ is the transverse momentum carried by particle
$i$ in the top jet $J_t$, and $\theta_i$ is the polar angle of particle $i$ with respect
to the top-jet axis.

\subsection{Top-quark Production and Decay}

\begin{center}
 \begin{figure}[htb]
  \def\SCALE{0.45}
  \def\OFFSET{20pt}
  \begin{tabular}{ccc}
   \includegraphics[scale=\SCALE]{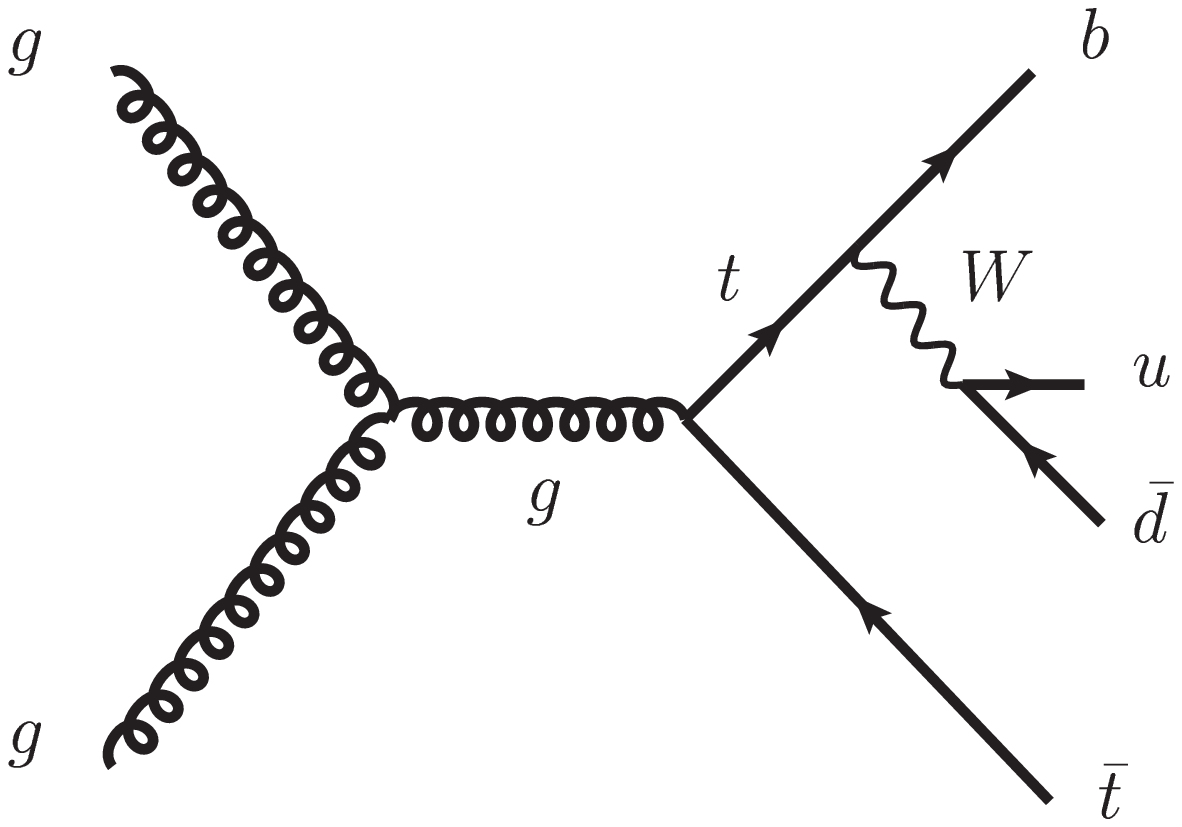} &
   \includegraphics[scale=\SCALE]{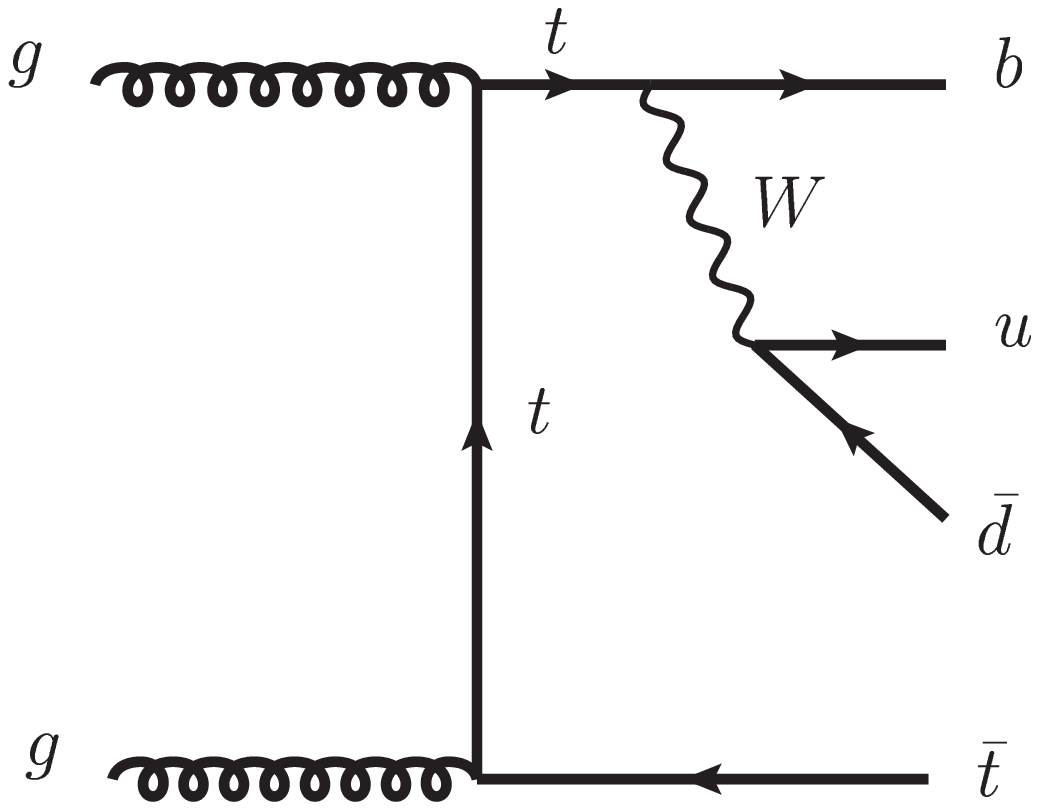} &
   \includegraphics[scale=\SCALE]{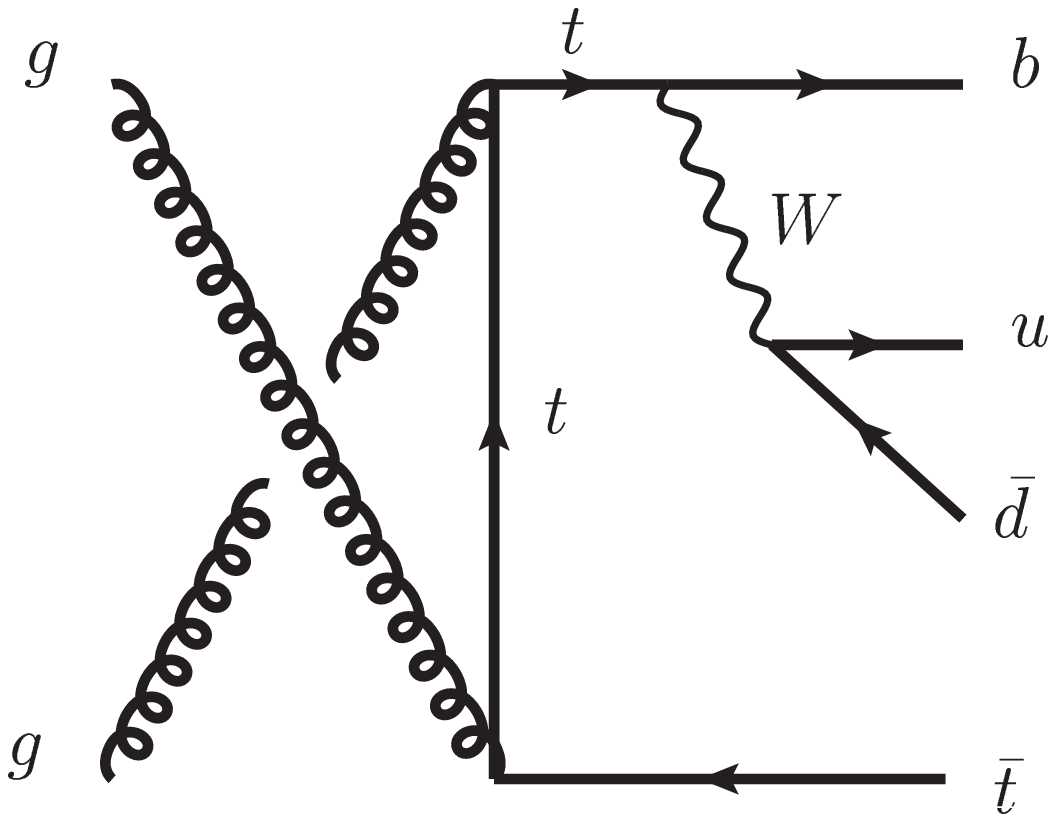}
  \end{tabular}
  \caption{LO diagrams for the top-pair production.} \label{fig-Feynman-diagrams}
 \end{figure}
\end{center}

We consider the
production process from the gluon fusion $gg\to t\bar{t}$ displayed in
Fig.~\ref{fig-Feynman-diagrams}, which dominates at the LHC.
The top quark then proceeds with the hadronic decay $t(k_t)\to b(k_b)u(k_u)\bar{d}(k_{\bar{d}})$,
where $k_i$ are the momenta of particle $i$, satisfying the conservation
$k_t = k_b + k_{u} + k_{\bar{d}}$.
Figure~\ref{fig-Feynman-diagrams} gives the squared amplitude
\begin{eqnarray}
 |\overline{\mathcal{M}}|^2
&=& \frac{g^4_s}{32^2} g^4 |V_{tb}|^2 |V_{ud}|^2 N_c D_{t}(k^2_t) D_{W}(k^2_W)
   \mbox{tr}\left[ k\sura_u \gamma^{\alpha}P_L k\sura_{\bar{d}}
	     \gamma^{\gamma}P_L \right]  \nn\\
&{}& \times
\sum_{i_b, i_{\bar{t}}}
   \mbox{tr}\left[ (k\sura_b + m_b) \gamma_{\alpha}P_L
                   (k\sura_{t} + m_t)
                   \Gamma^{ab,\mu\nu}_{i_b i_{\bar{t}}}
                   (k\sura_{\bar{t}} - m_t)
                  \bar{\Gamma}^{ab}_{i_b i_{\bar{t}},\mu\nu}
	          (k\sura_{t} + m_t)
                  \gamma_{\gamma}P_L \right],
\end{eqnarray}
in which $g_s$ is the QCD coupling, $g$ is the weak coupling, $V_{tb}$ and $V_{ud}$
are the Cabibbo-Kobayashi-Maskawa matrix element, $N_c=3$ is the number of colors,
$P_{L}=(1-\gamma^5)/2$ is the chiral projection matrix, and $m_{b}$ is the bottom
quark mass. The function
\begin{eqnarray}
D_{i}(k^2_{i})=\frac{1}{(k^2_i-m^2_i)^2+(m_i\Gamma_i)^2},
\end{eqnarray}
comes from the top quark ($W$-boson) propagator as $i=t$ $(i=W)$ with
the top-quark ($W$-boson) mass $m_t$ ($m_{W}$) and decay width $\Gamma_{t}$ ($\Gamma_{W}$).
The vertex function is written as
\begin{eqnarray}
\Gamma^{ab,\mu\nu}_{i_b i_{\bar{t}}}=
 - f^{abc}t^{c}_{i_b i_{\bar{t}}}\frac{1}{\hat{s}} \tilde{V}_{3}^{\mu\nu\rho}(k_1,k_2)
 + i t^{a}_{i_b k}t^{b}_{k i_{\bar{t}}} \gamma^{\mu}
   \frac{k\sura_t - k\sura_1 + m_t}{\hat{t} - m^2_t} \gamma^{\nu}
 + i t^{b}_{i_b k}t^{a}_{k i_{\bar{t}}} \gamma^{\nu}
   \frac{k\sura_t - k\sura_2 + m_t}{\hat{u} - m^2_t} \gamma^{\mu},\nn\\
\end{eqnarray}
with the three-gluon vertex
\begin{eqnarray}
 \tilde{V}^{\mu\nu\rho}_{3}(k_1,k_2)
&=&   g^{\mu\nu}(k_1 - k_2)^{\rho}
    + 2k_{2}^{\mu}g^{\nu\rho}
    - 2k_{1}^{\nu}g^{\rho\mu}.
\end{eqnarray}
In the above expressions
$i_{b,\bar{t}}$ label the colors of the $b,\bar{t}$ quarks,
$a,\mu,k_1$ and $b,\nu,k_2$ denote the colors, Lorentz indices, momenta of the incoming
gluons $1$ and $2$, respectively, and $ \hat{s} = (k_1 + k_2)^2$,  $\hat{t}  = (k_t - k_1)^2$, and
$\hat{u}= (k_t - k_2)^2$ are the kinematic invariants. Another vertex function
$\bar{\Gamma}^{ab,\mu\nu}_{i_b i_{\bar{t}}}$ is the Dirac conjugate of $\Gamma^{ab,\mu\nu}_{i_b i_{\bar{t}}}$,
defined as $ \bar{\Gamma}^{ab,\mu\nu}_{i_b i_{\bar{t}}}
 \equiv \gamma^{0} \left(\Gamma^{ab,\mu\nu}_{i_b i_{\bar{t}}}
                   \right)^{\dagger}\gamma^{0}$.

\begin{center}
 \begin{figure}[htb]
  \def\SCALE{0.5}
  \def\OFFSET{20pt}
  \begin{tabular}{ccc}
   \includegraphics[scale=\SCALE]{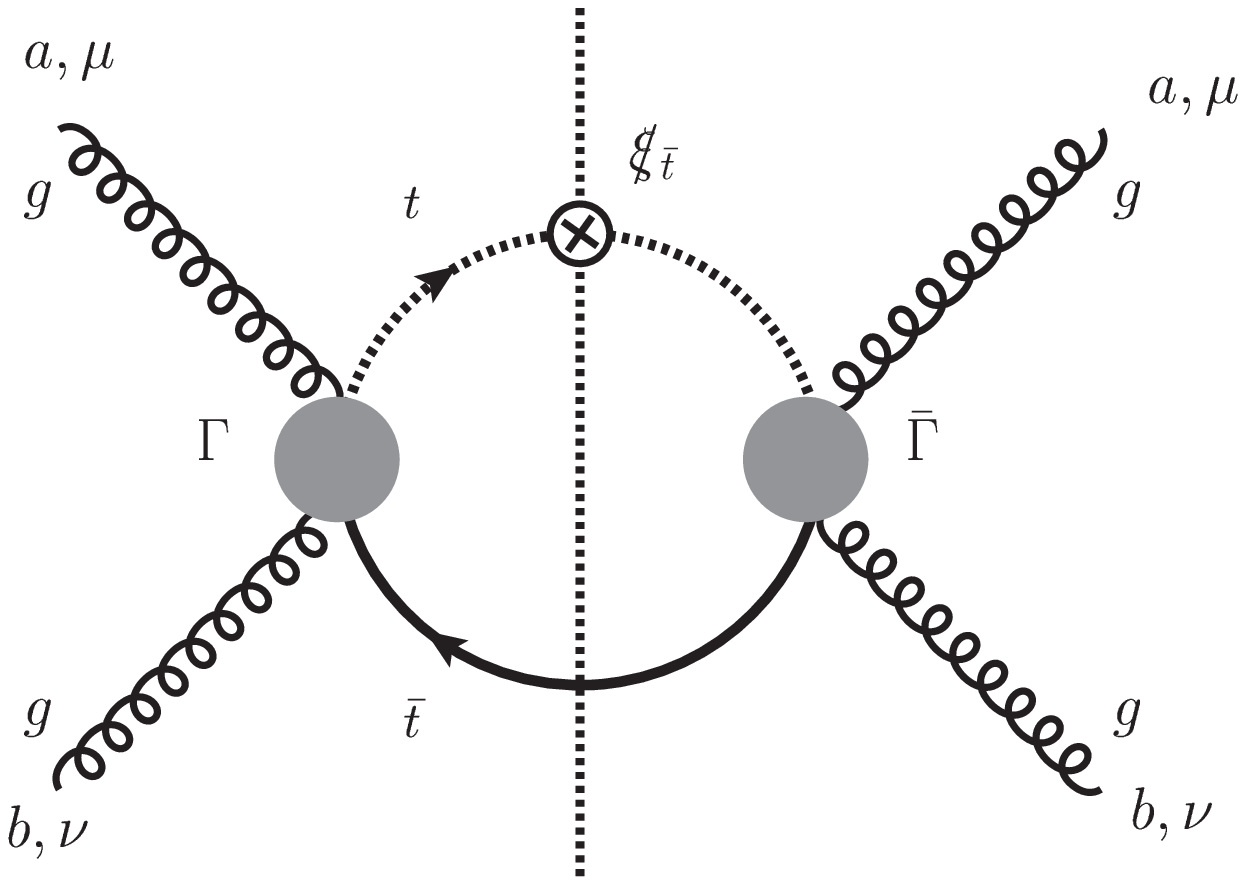} &
   \includegraphics[scale=\SCALE]{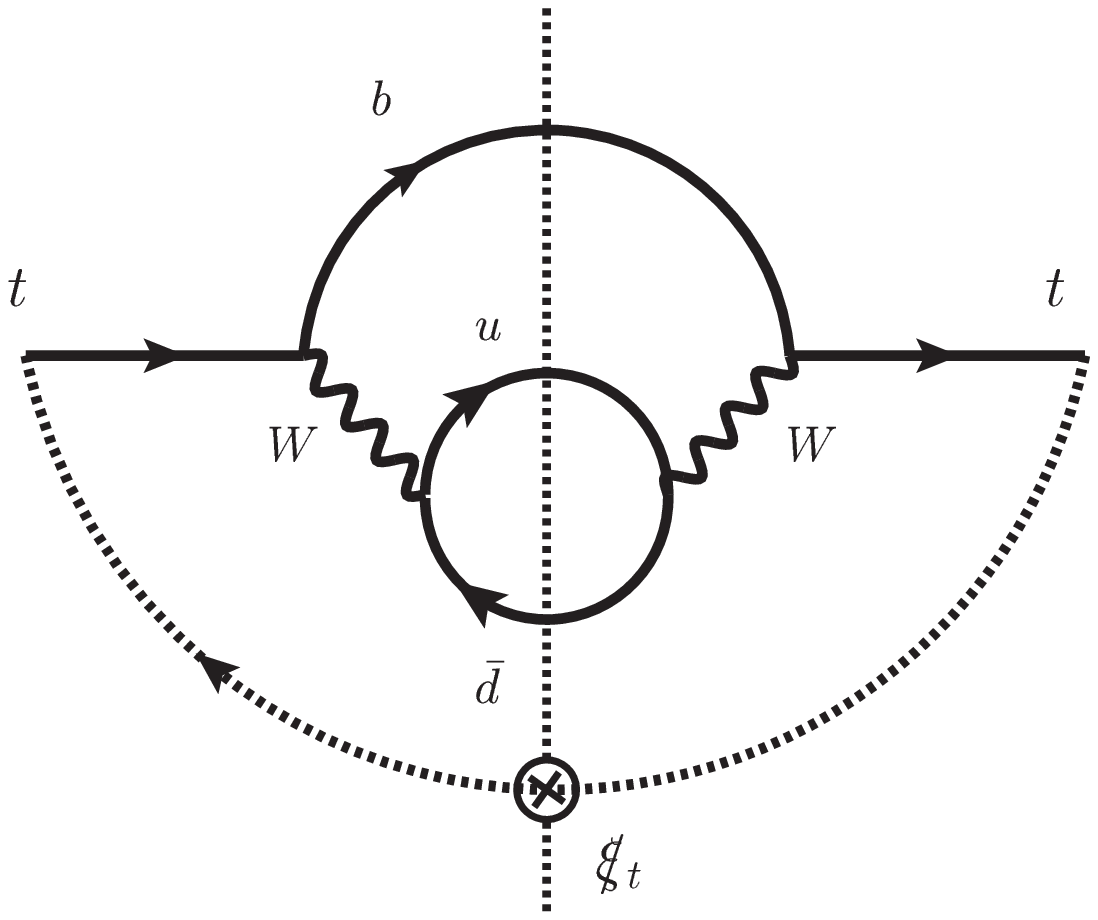} \\
   \hspace{-10pt} (a) &
   \hspace{30pt}  (b)
  \end{tabular}
  \caption{LO diagrams for (a) the production and (b) the hadronic decay of a top quark. The
dashed lines label the fermion flows, and the symbols $\otimes$ represent the insertions of the gamma
matrices $\bar{\xi}\sura_{t}$ and $\xi\sura_{t}$ from the Fierz identity.} \label{fig-trace-diagram}
 \end{figure}
\end{center}

The LO squared amplitude $ \Big| \overline{\mathcal{M}} \Big|^2$
is factorized, up to power corrections of ${\cal O}(m_t/E_t)$, $E_t$ being the
top-quark energy, into two pieces, i.e., the production part
$ \Big| \overline{\mathcal{M}}_{prod} \Big|^2 $ shown in Fig.~\ref{fig-trace-diagram}(a) and
the decay part $ \Big| \overline{\mathcal{M}}_{decay} \Big|^2 $ in
Fig.~\ref{fig-trace-diagram}(b):
\begin{eqnarray}
 |\overline{\mathcal{M}}|^2
&=&  |\overline{\mathcal{M}}_{prod}|^2
|\overline{\mathcal{M}}_{decay}|^2,\nn\\
|\overline{\mathcal{M}}_{prod}|^2
&\equiv&
 \frac{g^4_s}{16^2} \frac{1}{4}
   \sum_{i_{\bar{t}},i_b}\mbox{tr}\left[ \xi\sura_{\bar{t}}
                   \Gamma^{ab,\mu\nu}_{i_b i_{\bar{t}}}
                   (k\sura_{\bar{t}} - m_t)
                  \bar{\Gamma}^{ab}_{i_b i_{\bar{t}},\mu\nu}
            \right],\nn\\
|\overline{\mathcal{M}}_{decay}|^2
&=&  \frac{g^4}{4} N_c |V_{tb}|^2 |V_{ud}|^2 D_{t}(k^2_t) D_{W}(k^2_W)
\mbox{tr}\left[ k\sura_u \gamma^{\alpha}P_L k\sura_{\bar{d}}
	     \gamma^{\gamma}P_L \right] \nn\\
&{}& \times
   \mbox{tr}\left[ (k\sura_b + m_b) \gamma_{\alpha}P_L
                   (k\sura_{t} + m_t)
		   \xi\sura_{t}
	          (k\sura_{t} + m_t)
                  \gamma_{\gamma}P_L \right].
	     \label{eq.sqamp}
\end{eqnarray}
The dimesionless vectors \cite{ALP09}
\begin{eqnarray}
 \xi_{\bar{t}} &=&
  \frac{1}{\sqrt{2}}\left(1, -{\hat {n}}_{\bar{t}}\right),\hspace{1cm}
 \xi_{t} =
\frac{1}{\sqrt{2}}\left(1, -{\hat {n}}_{t}\right),
\end{eqnarray}
have been introduced via the Fierz transformation
to break the fermion flow \cite{Kitadono:2014hna}, in which
the unit vectors $\hat {n}_t$ and ${\hat {n}}_{\bar{t}}$ are along the directions of the
top-quark and anti-top-quark momenta, respectively.

\subsection{Top Jet Function}

The differential cross section for the $pp\to \bar{t}bu\bar{d}$ is then factorized into
the convolution of the production cross section $\sigma_{prod}$ with the top jet function $J_{t}$,
\begin{eqnarray}
 \frac{d\sigma(pp\to \bar{t}bu\bar{d})}{dE_{J_t} d\Omega_{J_t} dm^2_{J_t}}
 = \int dx_1 dx_2 \phi_g(x_1)\phi_g(x_2)\sigma_{prod}(x_1, x_2, m^2_{J_t}, E_{J_t}, \cos\theta_{J_t})
     J_{t}(m^2_{J_t}, E_{J_t}, R),\label{fact}
\end{eqnarray}
where $E_{J_t}$, $m_{J_t}$, $\Omega_{J_t}$, $\theta_{J_t}$, and $R$ are the energy,
the invariant mass, the solid angle, the polar angle, and the cone radius of the top
jet, respectively, and $\phi_g(x)$ is the gluon parton distribution function,
$x$ being the momentum fraction. The production cross section is written as
\begin{eqnarray}
\sigma_{prod}(x_1, x_2, m^2_{J_t}, E_{J_t}, \cos\theta_{J_t})
             &=& \frac{\pi \beta_{J_t}E_{J_t}}{4(2\pi)^3A_{t}\hat{s}E_{\bar{t}}}
                 \delta(E_1 + E_2 - E_{J_t} - E_{\bar{t}})
                 |\overline{\mathcal{M}}_{prod}|^2 ,
\end{eqnarray}
with the factor $\beta_{J_t}\equiv\sqrt{1-m^2_{J_t}/E^2_{J_t}}$, the gluonic parton
energies $E_1$ and $E_2$, and the anti-top quark energy $E_{\bar{t}}$.
The top jet function is defined as
\begin{eqnarray}
J_{t}(m^2_{J_t}, E_{J_t}, R)
&=& A_{t}  \int\frac{d^4k_{b}}{(2\pi)^3} \delta(k^2_{b}-m^2_b)
\int\frac{d^4k_{u}}{(2\pi)^3} \delta(k^2_{u})
     \int\frac{d^4k_{\bar{d}}}{(2\pi)^3} \delta(k^2_{\bar{d}}) |\overline{\mathcal{M}}_{decay}|^2\nn\\
& &\times \delta\left(m^2_{J_t}-(\sum k_{i})^2\right)\delta\left(E_{J_t}-\sum E_{i}\right)
\delta^{(2)}\left(\hat n_{J_t}-\frac{\sum {\vec k}_{i}}{|\sum {\vec k}_{i}|}\right),
\end{eqnarray}
in which the subscript $i$ runs over $b$, $u$ and $\bar{d}$, and
the three $\delta$-functions specify the top jet invariant mass,
energy and direction in the unit vector $\hat n_{J_t}$.
The normalization constant $A_t = (2\pi)^3/(2\sqrt{2}E^2_{J_t})$
has been chosen, such that the LO top jet function
reduces to $\delta(m^2_{J_t} - m^2_t)$ without the weak decay. It is
understood that the bottom, up and down quarks are restricted in the top jet cone.
At higher orders, QCD radiations from the final states in the top jet, that are
collimated to the top quark and in the jet cone, are grouped into $J_{t}$
straightforwardly. The collinear radiations from other subprocesses,
such as initial-state partons and the anti-top quark, are collected by the
Wilson lines in the direction of $\xi_{J_t}=(1, -{\hat {n}}_{J_t})/\sqrt{2}$
\cite{energyprofile}, which are associated with the definition of $J_{t}$.

The top jet function $J_{t}$ is first factorized into a hard top-quark decay kernel
$H_t$, a fat $W$-boson jet function $J_W$ with the cone radius $R$, and a fat bottom jet
function $J_{b}$ with the cone radius $R$, up to power corrections of
${\cal O}(m_{J_b}/m_{J_t})$,
\begin{eqnarray}
J_{t}(m^2_{J_t}, E_{J_t},R)
&=& \int dm^2_{J_W} dE_{J_W}d^2\hat n_{J_W}
    \int dm^2_{J_b} dE_{J_b}d^2\hat n_{J_b}H_{t} \nn\\
&{}&      \times J_{W}(m^2_{J_W}, E_{J_W}, R)J_{b}(m^2_{J_b}, E_{J_b}, R)
\nn\\
& &\times\delta\left(m^2_{J_t}-(k_{J_W}+k_{J_b})^2\right)
\delta\left(E_{J_t}-E_{J_W}-E_{J_b}\right)\nn\\
& &\times\delta^{(2)}\left(\hat n_{J_t}-\frac{{\vec k}_{J_W}+{\vec k}_{J_b}}
{|{\vec k}_{J_W}+{\vec k}_{J_b}|}\right).\label{top1}
\end{eqnarray}
The four-momentum, energy, and invariant mass of the $W$-boson (bottom) jet, 
$k_{J_W}$, $E_{J_W}$, and $m_{J_W}$ ($k_{J_b}$, $E_{J_b}$, and $m_{J_b}$), respectively,
are introduced according to the jet definitions \cite{Kitadono:2014hna}.
Below we will always work out the angular integration for the fat jet,
obtaining
\begin{eqnarray}
\int d^2\hat n_{J_W}\delta^{(2)}\left(\hat n_{J_t}-\frac{{\vec k}_{J_W}+{\vec k}_{J_b}}
{|{\vec k}_{J_W}+{\vec k}_{J_b}|}\right)=\frac{|{\vec k}_{J_W}+{\vec k}_{J_b}|}{|{\vec k}_{J_W}|}
=\frac{\beta_{J_t}E_{J_t}}{\beta_{J_W}E_{J_W}}.
\end{eqnarray}

The bottom jet will be treated as a light-quark jet here as stated before.
The integration of $|\overline{\mathcal{M}}_{decay}|^2$, which
contains the $W$-boson propagator, over the $u,\bar{d}$ kinematic variables is
proportional to the sum over the $W$-boson polarizations,
$d^{\alpha\gamma}=g^{\alpha\gamma}-k^{\alpha}_{J_W}k^{\gamma}_{J_W}/m^2_{J_W}$.
Therefore, we extract the $W$-boson jet function according to
\begin{eqnarray}
 -d^{\alpha\gamma} J_{W}(m^2_{J_W}, E_{J_W}, R) &=&
 g^2|V_{ud}|^2N_cA_{W}\int \frac{d^4k_{u}}{(2\pi)^3} \delta(k^2_{u})
     \int\frac{d^4k_{\bar{d}}}{(2\pi)^3} \delta(k^2_{\bar{d}})\nn\\
& &\times \mbox{tr} \left[ k\sura_{u} \gamma^{\alpha}P_{L} k\sura_{\bar{d}}\gamma^{\gamma}P_{L}
               \right] D_{W}(k_{J_W}^2)\delta\left(m^2_{J_W}-(k_{u}+k_{{\bar d}})^2\right)
     \nn\\
& &\times
\delta\left(E_{J_W}-E_{u}-E_{{\bar d}}\right)
\delta^{(2)}\left(\hat n_{J_W}-\frac{{\vec k}_{u}+{\vec k}_{\bar d}}
{|{\vec k}_{u}+{\vec k}_{\bar d}|}\right),
\end{eqnarray}
where the normalization constant
$A_{W}=2(2\pi)^3/E_{J_W}$ is chosen, such that the LO $W$-boson jet function without
the weak decay reduces to $\delta(m^2_{J_W}-m^2_{W})$.  At higher orders, $J_W$
absorbs the collinear radiations collimated to the $W$-boson and the soft radiations
in the top jet.
The LO hard kernel $H_{t}$ is given by
\begin{eqnarray}
 H_{t}
&=& \frac{g^2}{8\sqrt{2}}|V_{tb}|^2\frac{A_{t}E_{J_t}}{A_{W}A_{b}}
    (1+\beta_{J_t}) D_{t}(k_{J_t}^2)
      (-d^{\alpha\gamma}) \mbox{tr} \left[ \gamma_{\alpha}P_L
                   (k\sura_{J_t} + m_t)
                    \gamma_{\gamma}P_L \bar{\xi}\sura_{J_b} \right] \nn\\
&& + {\cal O}(k^2_{J_t} - m^2_t), \label{eq.def.Ht}
\end{eqnarray}
where the normalization constant $A_{b}=(2\pi)^3/(2\sqrt{2}E^2_{J_b})$
is introduced through the definition of $J_b$ \cite{Kitadono:2014hna},
and the light-like vector $\bar\xi_{J_b}$ is along the direction of the
bottom jet momentum.

Next we factorize the fat $W$-boson jet function into
a hard $W$-boson decay kernel, a fat up jet function with the cone radius $R$,
and a thin down jet function with the small cone radius $r$.
Using the relation $g_{\alpha\gamma}d^{\alpha\gamma}=3$, we write $J_{W}$ as
\begin{eqnarray}
 J_{W}(m^2_{J_W}, E_{J_W},R)
&=& \int dm^2_{J_u} dE_{J_u}
    \int dm^2_{J_{\bar{d}}} dE_{J_{\bar{d}}} d^2\hat n_{J_{\bar{d}}}
    \frac{\beta_{J_W}E_{J_W}}{\beta_{J_u}E_{J_u}}H_{W}\nn\\
&{}&      \times   J_{u}(m^2_{J_u}, E_{J_u}, R)
J_{\bar{d}}(m^2_{J_{\bar{d}}}, E_{J_{\bar{d}}}, r)\nn\\
& &\times\delta\left(m^2_{J_W}-(k_{J_u}+k_{J_{\bar{d}}})^2\right)
\delta\left(E_{J_W}-E_{J_u}-E_{J_{\bar{d}}}\right)
                        ,\label{Wjet}
\end{eqnarray}
with the LO hard kernel
\begin{eqnarray}
 H_{W}&=&g^2|V_{ud}|^2N_c\frac{A_{W}}{A_uA_{\bar d}}
\frac{-g_{\alpha\gamma}}{3}\left[ \bar{\xi}\sura_{J_u} \gamma^{\alpha}P_{L}
\bar{\xi}\sura_{J_{\bar{d}}}\gamma^{\gamma}P_{L}
\right] D_{W}(k_{J_W}^2),\nn\\
&=&  \frac{g^2}{12\pi^3}|V_{ud}|^2N_c\frac{E_{J_u}^2E_{J_{\bar d}}^2}{E_{J_W}}
[1-\cos(\theta_{J_u}+\theta_{J_{\bar{d}}})]
     D_{W}(k^2_{J_W}).\label{hw}
\end{eqnarray}
In the above expressions $k_{J_u}$, $E_{J_u}$, $\theta_{J_u}$, and $m_{J_u}$ ($k_{J_{\bar d}}$,
$E_{J_{\bar d}}$, $\theta_{J_{\bar d}}$, and $m_{J_{\bar d}}$) are the four-momentum, energy, angle,
and invariant mass of the up-quark (down-quark) jet, respectively. 
The constant $A_u=(2\pi)^3/(2\sqrt{2}E^2_{J_u})$
($A_{\bar d}=(2\pi)^3/(2\sqrt{2}E^2_{J_d})$) is introduced to normalize the up (down)
jet function to a $\delta$-function at LO in $\alpha_s$ \cite{energyprofile}.
To factorize the up and down jet functions from the $W$-boson jet function,
we also need to introduce the light-like vectors $\bar{\xi}_{J_u}$ and $\bar{\xi}_{J_{\bar{d}}}$
along the directions of the up-jet and down-jet momenta,
respectively, as a consequence of the Fierz transformation \cite{energyprofile}. Their
inner product gives $\bar{\xi}_{J_u}\cdot\bar{\xi}_{J_{\bar{d}}}=
[1-\cos(\theta_{J_u}+\theta_{J_{\bar{d}}})]/2$ in the second line of Eq.~(\ref{hw}).

At last, to discuss the dependence on the top-quark helicity,
we insert the top-spin projectors
\begin{eqnarray}
 w_{s_t} &=& \frac{1}{2}({\bf 1} + \gamma^5s\sura_t), \hspace{1cm}
 \bar{w}_{s_t} = \frac{1}{2}({\bf 1} - \gamma^5 s\sura_t),
\end{eqnarray}
under which the unpolarized top jet function is decomposed into
$J_{t} = J^{s_t}_{t} + J^{\bar{s}_t}_{t}$.
The corresponding hard top-quark decay kernel for $J^{s_t}_{t}$ is given by
\begin{eqnarray}
 H^{s_t}_{t}
 &=&   \frac{g^2}{8\sqrt{2}}|V_{tb}|^2\frac{A_{t}E_{J_t}}{A_{W}A_{b}}
 (1+\beta_{J_t}) D_{t}(k_{J_t}^2)
      (-d^{\alpha\gamma}) \mbox{tr} \left[ \gamma_{\alpha}P_L
                   (k\sura_{J_t} + m_t)w_{s_t}
                    \gamma_{\gamma}P_L \bar{\xi}\sura_{J_b} \right],\nn\\
&=& \frac{g^2}{16(2\pi)^3}|V_{tb}|^2\frac{E_{J_W}E_{J_b}}{E_{J_t}}
(1+\beta_{J_t})D_{t}(k_{J_t}^2)\nn\\
&{}& \times
     \left[  \left( \frac{m^2_{J_t}}{m^2_{J_W}} +2\right)
             \left( \frac{m^2_{J_t} - m^2_{J_W}}{2} \right)
           + \left( \frac{m^2_{J_t}}{m^2_{J_W}}-2 \right)
	      m_t E_{J_b}
             \left( s^{0}_t - |\vec{s}_t|\cos\theta_{J_b} \right)
     \right],\label{hst}
\end{eqnarray}
and $H^{\bar{s}_t}_{t}$ for $J^{\bar{s}_t}_{t}$ is similar, but with $w_{s_t}$ being
replaced by $\bar{w}_{s_t}$. The $W$-boson jet function and the bottom jet function
in the factorization formulas for $J^{s_t}_{t}$ and $J^{\bar{s}_t}_{t}$ are the same as
for the unpolarized top jet function. The absolute value of the top
spin in the boosted frame is set to $|\vec{s}_t|=1/\sqrt{1-\beta_{J_t}^2}$, and
the zeroth component is then given by $s^{0}_{t}=\beta_{J_t}|\vec{s}_t|$.

\subsection{Top Jet Energy Function}

Following the reasoning in \cite{energyprofile}, we construct the polarized
top jet energy function $J^{E,s_t}_{t}$ by accumulating the energy of
final-state particles in the test cone of radius $r$.
It is straightforward to derive the factorization formula
\begin{eqnarray}
 J^{E,s_t}_{t}(m^2_{J_t}, E_{J_t}, R, r)
 &=& \int dm^2_{J_W} dE_{J_W}
 \int dm^2_{J_b}dE_{J_b}d^2\hat n_{J_b}
 \frac{\beta_{J_t}E_{J_t}}{\beta_{J_W}E_{J_W}}H^{s_t}_{t}  \nn\\
&{}& \times
 \left[ J^{E}_{b}(m^2_{J_b}, E_{J_b}, R, r) J_{W}(m^2_{J_W}, E_{J_W}, R) \Theta(ar-\theta_{J_b})
\right.\nn\\
&{}& \left. \hspace{0.5cm}
       + J^{E}_{W}(m^2_{J_W}, E_{J_W}, R, r) J_{b}(m^2_{J_b}, E_{J_b}, R) \Theta(ar-\theta_{J_W})
 \right]\nn\\
& &\times \delta\left(m^2_{J_t}-(k_{J_W}+k_{J_b})^2\right)
\delta\left(E_{J_t}-E_{J_W}-E_{J_b}\right)
, \label{tjep}
\end{eqnarray}
where $a\sim {\cal O}(1)$ is a geometric factor to be specified below.
The bottom ($W$-boson) jet energy function $J^{E}_{b(W)}$ absorbs QCD radiations,
which are associated with the bottom ($W$-boson) jet and go into the test cone.
That is, the first (second) term describes the contribution to the
top jet energy profile from the bottom ($W$-boson) jet.
The $\Theta$ function requires the polar angle of a subjet to
be smaller than $ar$, if it contributes to the energy profile.

It should be pointed out that the contribution to the energy profile from the
hard gluons has been neglected in Eq.~(\ref{tjep}), which 
appears at next-to-leading order in QCD. To be consistent with this accuracy, the
parameter $a$ is set to $a=2$ to minimize the contribution
from hard gluons \cite{boost.higgs.subjet}. In the present work we will
approximate the fat bottom jet by a thin bottom jet of cone radius $r$,
for which the simplification $J^{E}_{b}(m^2_{J_b}, E_{J_b}, r, r)
= E_{J_b} J_{b}(m^2_{J_b}, E_{J_b}, r)$ holds. 
The overestimate of the top jet energy profile due to this approximation can be
compensated by choosing a value of $a<2$. We have confirmed that $a=1.7$
is the best choice, which reproduces the fat bottom jet contribution
in the considered kinematic region. In other words, the choice $a=1.7$ 
includes the resummation effect in the fat jet contribution. 
Neglecting the small bottom jet
invariant mass $m_{J_b}$ in the hard kernel $H^{s_t}_{t}$ and in the
$\delta$-functions, we can integrate out the $m_{J_b}$ dependence, and have
$\int dm^2_{J_b} J_{b}(m^2_{J_b}, E_{J_b}, r) = 1 + \mathcal{O}(\alpha_s)$.

The narrow-width approximation is then applied to both the $W$-boson and top-quark
propagators $D_{i=t,W}$, 
\begin{eqnarray}
D_{i}(m_{J_i}^2) \approx \frac{\pi}{m_{i}\Gamma_{i}} \delta(m^2_{J_i} - m^2_{i}),
\label{narrow}
\end{eqnarray}
so that the integration over the $W$-boson jet invariant mass $m^2_{J_W}$
in Eq.~({\ref{tjep}) can be carried out trivially, with all $m^2_{J_W}$ being
replaced by the $W$-boson mass $m^2_{W}$.
The polarized top jet energy function is then simplified into
\begin{eqnarray}
 J^{E,s_t}_{t}(m^2_{J_t}, E_{J_t}, R, r)
 &=&
\int dE_{J_W}\int dE_{J_b}d^2\hat n_{J_b}
 \frac{\beta_{J_t}}{\beta_{W}}R^2E_{J_t}E_{J_W}H^{s_t}_{t}  \nn\\
&{}& \times
 \left[ E_{J_b}\bar{J}_{W}(N=1, E_{J_W}, R) \Theta(ar-\theta_{J_b})
\right.\nn\\
&{}& \left. \hspace{0.5cm}
       + \bar{J}^{E}_{W}(N=1, E_{J_W}, R, r)  \Theta(ar-\theta_{J_W})
 \right]\nn\\
& &\times \delta\left(m^2_{J_t}-(k_{J_W}+k_{J_b})^2\right)\delta\left(E_{J_t}-E_{J_W}-E_{J_b}\right)
 ,\label{tjep1}
\end{eqnarray}
with the factor $\beta_{W}=\sqrt{1-m_W^2/E_{J_W}^2}$, and the first moments of the 
$W$-boson jet function and energy function
\begin{eqnarray}
 \int \frac{dm^2_{J_W}}{(R E_{J_W})^2} J_{W}(m^2_{J_W}, E_{J_W}, R)
 &=& \bar{J}_{W}(N=1,E_{J_W},R),\nn\\
\int \frac{dm^2_{J_W}}{(R E_{J_W})^2} J^{E}_{W}(m^2_{J_W}, E_{J_W}, R, r)
 &=& \bar{J}^{E}_{W}(N=1,E_{J_W},R,r),\label{bn}
\end{eqnarray}
respectively.

As in \cite{Kitadono:2014hna}, we boost the top quark such that $H^{s_t}_{t}$ in
Eq.~(\ref{hst}) corresponds to the top-quark decay kernel of helicity plus $h=+$ and
$H^{\bar{s}_t}_{t}$ corresponds to helicity minus $h=-$.
The narrow-width approximation for the top-quark propagator allows us to
compute the first moment of the top jet energy function $J^{E,h=\pm}_{t}$ easily,
\begin{eqnarray}
\bar{J}^{E,h=\pm}_{t}(1,E_{J_t},R,r)&=&
\frac{g^2}{64\pi}|V_{tb}|^2(1+\beta_t)\beta_t
\frac{E_{J_t}^2}{m_t\Gamma_{t}}\int dz_{J_b}d\cos\theta_{J_b} z_{J_b}(1-z_{J_b})^2 \nn\\
&{}& \times
     \left[  \left( \frac{m^2_{t}}{m^2_{W}} +2\right)
             \left( \frac{m^2_{t} - m^2_{W}}{2} \right)\right.\nn\\
& &\left.  \hspace{0.5cm}         \pm \left( \frac{m^2_{t}}{m^2_{W}}-2 \right)
	      z_{J_b}m_t E_{J_t}
             \left( s^{0}_t - |\vec{s}_t|\cos\theta_{J_b} \right)
     \right],\nn\\
&{}& \times\frac{1}{\beta_W}
 \left[ z_{J_b}E_{J_t}\bar{J}_{W}(1, (1-z_{J_b})E_{J_t}, R) \Theta(ar-\theta_{J_b})\right.\nn\\
 & &\left.\hspace{1.0cm}
       + \bar{J}^{E}_{W}(1, (1-z_{J_b})E_{J_t}, R, r)  \Theta(ar-\theta_{J_W})
 \right]\nn\\
& &\times \delta\left(m^2_{t}-m^2_W-2z_{J_b}
(1-z_{J_b})E_{J_t}^2[1-\beta_W\cos(\theta_{J_W}+\theta_{J_b})]\right),\label{tjep2}
\end{eqnarray}
with the factor $\beta_{t}=\sqrt{1-m_t^2/E_{J_t}^2}$ and
the energy fraction $z_{J_b}=E_{J_b}/E_{J_t}$ for the bottom jet.

The factorization formula for the $W$-boson jet energy function
is written as
\begin{eqnarray}
 J^E_{W}(m^2_{J_W}, E_{J_W},R,r)
&=& \int dm^2_{J_u} dE_{J_u}
    \int dm^2_{J_{\bar{d}}} dE_{J_{\bar{d}}} d^2\hat n_{J_{\bar{d}}}
    \frac{\beta_{J_W}E_{J_W}}{\beta_{J_u}E_{J_u}}H_{W} \nn\\
&{}&      \times  \left[J^E_{u}(m^2_{J_u}, E_{J_u}, R,r)
J_{\bar{d}}(m^2_{J_{\bar{d}}}, E_{J_{\bar{d}}}, r)\Theta(ar-\theta_{J_u})\right.\nn\\
& &\left.+ J_{u}(m^2_{J_u}, E_{J_u}, R)J^E_{\bar{d}}(m^2_{J_{\bar{d}}}, E_{J_{\bar{d}}}, r,r)
\Theta(ar-\theta_{J_{\bar{d}}})\right]\nn\\
& &\times \delta\left(m^2_{J_W}-(k_{J_u}+k_{J_{\bar{d}}})^2\right)
\delta\left(E_{J_W}-E_{J_u}-E_{J_{\bar{d}}}\right),
\label{wjep1}\\
&\approx& 2\int dm^2_{J_u} dE_{J_u}
    \int dm^2_{J_{\bar{d}}}dE_{J_{\bar{d}}} d^2\hat n_{J_{\bar{d}}}
    \frac{\beta_{J_W}E_{J_W}}{\beta_{J_u}E_{J_u}}H_{W}\nn\\
&{}&      \times
J_{u}(m^2_{J_u}, E_{J_u}, R)J^E_{\bar{d}}(m^2_{J_{\bar{d}}}, E_{J_{\bar{d}}}, r,r)
\Theta(ar-\theta_{J_{\bar{d}}})\nn\\
& &\times \delta\left(m^2_{J_W}-(k_{J_u}+k_{J_{\bar{d}}})^2\right)
\delta\left(E_{J_W}-E_{J_u}-E_{J_{\bar{d}}}\right),
\label{wjep2}
\end{eqnarray}
where the meaning of each factor can be understood in a similar way.
Inputting the jet functions and the jet energy functions from the resummation formalism
\cite{energyprofile}, one can compute $J^E_{W}(m^2_{J_W}, E_{J_W},R,r)$ in principle.
Here we have approximated the fat jet contribution to the $W$-boson jet energy function
by the thin jet contribution again with the parameter $a=1.7$, as deriving Eq.~(\ref{wjep2})
from Eq.~(\ref{wjep1}) .
 
To get analytical expressions for the $W$-boson jet function and energy function, 
we simplify the thin down jet energy function into
$ J^{E}_{\bar{d}}(m^2_{J_{\bar{d}}}, E_{J_{\bar{d}}}, r, r)
 = E_{J_{\bar{d}}} J_{\bar{d}}(m^2_{J_{\bar{d}}}, E_{J_{\bar{d}}}, r)$,
and approximate the up jet function by a $\delta$-function \cite{boost.higgs.subjet}
and $\beta_{J_u}$ by $\beta_{J_u}\approx 1$ in
Eq.~(\ref{wjep2}). The conservation of the
momenta perpendicular to the $W$-boson jet axis gives the relation
$E_{J_u}\theta_{J_u}=E_{J_{\bar d}}\theta_{J_{\bar d}}$
in the small angle limit, i.e., in the small $m_W/E_{J_W}$ limit.
The narrow-width approximation then determines the relation
between the down jet energy fraction $y_{J_{\bar d}}=E_{J_{\bar d}}/E_{J_W}$ and the
polar angle $\theta_{J_{\bar d}}$,
\begin{eqnarray}
\theta_{J_{\bar d}}^2\approx\frac{E_{J_u}m_W^2}{E_{J_{\bar d}}E_{J_W}^2}.
\label{thetad}
\end{eqnarray}
The $W$-boson jet function and energy function reduce to
\begin{eqnarray}
{\bar J}_{W}(N=1,E_{J_W},R)& =&
\frac{g^2}{4\pi}
\frac{\beta_{W}m_{W}}{R^2E_{J_W}^2\Gamma_W }|V_{ud}|^2
\int_{y^m}^{y^M}dy_{J_{\bar d}} (1-y_{J_{\bar d}}),\label{wjf}\\
{\bar J}_{W}^E(N=1,E_{J_W},R,r)& =&\frac{g^2}{2\pi}
\frac{\beta_{W}m_{W}}{R^2E_{J_W}\Gamma_W }|V_{ud}|^2
\int_{y_E^m}^{y_E^M}dy_{J_{\bar d}} y_{J_{\bar d}}(1-y_{J_{\bar d}}),\label{wjep}
\end{eqnarray}
where $N_c=3$ has been adopted in Eq.~(\ref{hw}).

The requirement that both light-quark jets are inside the $W$-boson jet cone give
the constraints $\theta_{J_{\bar d}}^2=(1-y_{J_{\bar d}}){\hat m}_W^2/y_{J_{\bar d}} < R^2$ and
$\theta_{J_u}^2=y_{J_{\bar d}}{\hat m}_W^2/(1-y_{J_{\bar d}}) < R^2$ for Eq.~(\ref{Wjet})
with $\hat{m}_{W} = m_{W}/E_{J_W}$. They lead to the integration range
$y^m< y_{J_{\bar d}} < y^M$ in Eq.~(\ref{wjf}) with the bounds
\begin{eqnarray}
   y^m=\frac{\hat{m}^2_{W}}{R^2+\hat{m}^2_{W}},\hspace{1cm}
   y^M=\frac{R^2}{R^2+\hat{m}^2_{W}}.\label{yb}
\end{eqnarray}
The above inequality implies that the contribution from the region of $E_{J_W}\to m_W$ is suppressed by
phase space. The similar requirement together with the $\Theta$ function in Eq.~(\ref{wjep2})
give $(1-y_{J_{\bar d}}){\hat m}_W^2/y_{J_{\bar d}} < \min(R^2,a^2r^2)$ and
$y_{J_{\bar d}}{\hat m}_W^2/(1-y_{J_{\bar d}}) < R^2$,
which set the bounds in Eq.~(\ref{wjep}),
\begin{eqnarray}
   y_E^m=\frac{\hat{m}^2_{W}}{\hat{m}^2_{W} + \min(R^2,a^2r^2)},\hspace{1cm}
   y_E^M=\frac{R^2}{R^2+\hat{m}^2_{W}}.\label{yeb}
\end{eqnarray}

The $\delta$-function in Eq.~(\ref{tjep2}) leads, in the small angle limit, i.e.,
in the expansion of $m_W/E_{J_W}$ and $m_t/E_{J_t}$, to
\begin{eqnarray}
\theta_{J_b}^2&\approx&\frac{E_{J_W}m_t^2-E_{J_t}m_W^2}{E_{J_b}E_{J_t}^2},\nn\\
\theta_{J_W}^2&\approx&\frac{E_{J_b}(E_{J_W}m_t^2-E_{J_t}m_W^2)}{E_{J_W}^2E_{J_t}^2}.
\label{thetab}
\end{eqnarray}
Equation~(\ref{tjep2}) is then simplified into
\begin{eqnarray}
\bar{J}^{E,h=\pm}_{t}(1,E_{J_t},R,r)&=&
\frac{g^4}{256\pi^2}\frac{|V_{tb}|^2|V_{ud}|^2m_W}{R^2\Gamma_{t}\Gamma_{W}}
\int dz_{J_b} (1-z_{J_b}) \nn\\
&{}& \times
     \left[  \left( \frac{m^2_{t}}{m^2_{W}}+2 \right)
             \left( \frac{m^2_{t} - m^2_{W}}{2m_tE_{J_t}} \right)
             \pm z_{J_b}\left( \frac{m^2_{t}}{m^2_{W}}-2 \right)
             \left( s^{0}_t - |\vec{s}_t|\cos\theta_{J_b} \right)
     \right],\nn\\
&{}& \times
 \left[ z_{J_b}\int_{y^m}^{y^M}dy_{J_{\bar d}} (1-y_{J_{\bar d}})
 \Theta(ar-\theta_{J_b})\right.\nn\\
 & &\left.\hspace{0.5cm}
       + 2(1-z_{J_b})\int_{y_E^m}^{y_E^M}dy_{J_{\bar d}} y_{J_{\bar d}}(1-y_{J_{\bar d}})
       \Theta(ar-\theta_{J_W})
 \right].\label{tjep3}
\end{eqnarray}

The two step functions $\Theta(ar-\theta_{J_b})$ and $\Theta(ar-\theta_{J_W})$
constrain the integration variable $z_{J_b}$ for the bottom jet and
$W$-boson jet contributions, respectively. Together with Eq.~(\ref{yb}), which requires
$\hat{m}_{W}<R$, we derive the bounds $z_1^m<z_{J_b}<z_1^M$ for the former
with
\begin{eqnarray}
   z_1^m=\frac{\tilde{m}^2_{t}-\tilde{m}^2_{W}}{\tilde{m}^2_{t} + \min(R^2,a^2r^2)},\hspace{1cm}
   z_1^M=1-\frac{\tilde{m}_{W}}{R},
\end{eqnarray}
and $\tilde{m}_{t}=m_{t}/E_{J_t}$ and $\tilde{m}_{W}=m_{W}/E_{J_t}$.
It is trivial to verify that the above lower and upper bounds always obey their
inequality for arbitrary $r$, $R\sim {\cal O}(1)$ and $E_{J_t}\sim
{\cal O}(1)$ TeV. That is, there is no significant dead-cone effect for the bottom jet
contribution. Besides, the $W$-boson jet always remains inside the top jet cone
of radius $R$ in the above range of $z_{J_b}$. 

As to the $W$-boson jet contribution,
we have $z_{J_b}<z_2^M$ for the lower and upper 
bounds in Eq.~(\ref{yeb}) to hold their inequality. The requirement that the bottom jet
is inside the top jet cone sets $z_{J_b}> z_2^m$. The combination
of the above two constraints,$z_2^m < z_{J_b} < z_2^M$, with
\begin{eqnarray}
z_2^m=\frac{\tilde{m}^2_{t}-\tilde{m}^2_{W}}{R^2+\tilde{m}^2_{t} },\hspace{1cm}
z_2^M=1-\frac{\tilde{m}_{W}}{[R^2\min(R^2,a^2r^2)]^{1/4}},\label{2nd}
\end{eqnarray}
implies
\begin{eqnarray}
ar > \left(\frac{R^2+\tilde{m}_{t}^2}{R^2+\tilde{m}_{W}^2}\right)^2\frac{\tilde{m}_{W}^2}{R}, \label{eq.rmin}
\end{eqnarray}
namely, a more significant dead-cone effect for the $W$-boson jet contribution.
For $ar< (\tilde{m}^2_{t}-\tilde{m}^2_{W})/(2\tilde{m}_{W})$, there exist additional constraints
from the $W$-boson jet to be inside the test cone or the top jet cone, $z_{J_b}<z_2^{m\prime}$
or $z_{J_b}>z_2^{M\prime}$, with
\begin{eqnarray}
 z_2^{m\prime}&=&\frac{2\min(R^2,a^2r^2)+\tilde{m}^2_{t}-\tilde{m}^2_{W}
   -\sqrt{(\tilde{m}^2_{t}-\tilde{m}^2_{W})^2-4\min(R^2,a^2r^2)\tilde{m}^2_{W}}}
   {2( R^2+\tilde{m}^2_{t})},\nn\\
 z_2^{M\prime}&=&\frac{2\min(R^2,a^2r^2)+\tilde{m}^2_{t}-\tilde{m}^2_{W}
   +\sqrt{(\tilde{m}^2_{t}-\tilde{m}^2_{W})^2-4\min(R^2,a^2r^2)\tilde{m}^2_{W}}}
   {2( R^2+\tilde{m}^2_{t})}.\label{add}
\end{eqnarray}
The allowed region of $z_{J_b}$ for the $W$-boson jet contribution in Eq.~(\ref{tjep3})
then comes from the overlap of the above range with $z_2^m < z_{J_b} < z_2^M$. Note 
that the bounds in Eq.~(\ref{add}) are effective only
at small $r\sim 0.1$ for the top jet energy $E_{J_t}> 500$ GeV.

\section{Numerical Analysis \label{result}}

In this section we evaluate the jet energy profiles of a boosted polarized
hadronic top quark derived in Eq.~(\ref{tjep3}).
In the moment space the convolution in the momentum fractions in Eq.~(\ref{fact})
becomes a product, such that the top jet energy profile $\Psi(E_{J_t},R,r)$
can be simply calculated as the ratio
\begin{eqnarray}
 \Psi^{\pm}(E_{J_t}, R, r) &=& \frac{\bar{J}^{E,h=\pm}_{t}(1,E_{J_t},R,r) }
 {\bar{J}^{E,h=\pm}_{t}(1,E_{J_t},R,R) },
\end{eqnarray}
in which the production part has been canceled between the numerator and the
denominator. We take the parameters
$m_{t}=173.5$ GeV, $m_{W}=80.39$ GeV, $\Gamma_{W}=2.09$ GeV, $|V_{tb}|=1.0$,
and $R=0.7$ for the numerical analysis below.

We display the energy profiles of the top jet, which come only from the bottom-jet contribution,
in Figs.~\ref{fig-psi-b}(a), \ref{fig-psi-b}(b) and, \ref{fig-psi-b}(c) for the top jet energy
$E_{J_t}=500$ GeV, 1 TeV and 2 TeV, respectively.
It is found that the bottom jet contributes more to the energy profile of
a helicity-minus top quark, similar to the semi-leptonic top case \cite{Kitadono:2014hna}.
The difference of the energy profiles between two opposite helicities decreases, when the top
jet energy increases as expected \cite{Kitadono:2014hna}.
The top jet energy profiles, which contain only the $W$-boson jet contribution,
are shown in Figs.~\ref{fig-psi-w}(a), \ref{fig-psi-w}(b), and \ref{fig-psi-w}(c) for
$E_{J_t}=500$ GeV, 1 TeV and 2 TeV, respectively. Contrary to the bottom jet,
the $W$-boson jet contributes more to the energy profile of a right-handed top quark.
It has been explained in \cite{Kitadono:2014hna} that the above results are attributed
to the feature of the $V-A$ weak interaction. Moreover, the $W$-boson contribution
exhibits a more obvious dead-cone effect as highlighted by the sharp increase
of the curves with the test cone radius $r$ in Fig.~\ref{fig-psi-w}(a).

The jet energy profiles of a boosted hadronic top quark including
the contributions from both the bottom and $W$-boson jets are exhibited
in Figs.~\ref{fig-psi-tot}(a), \ref{fig-psi-tot}(b), and
\ref{fig-psi-tot}(c) for $E_{J_t}=500$ GeV, 1 TeV and 2 TeV, respectively.
Since the bottom jet contributes more to the left-handed top, and
the $W$-boson jet contributes more to the right-handed top, it turns out that
their contributions compensate each other, and there is tiny dependence
of the energy profiles on the top-quark helicity. On the other hand, the energy
profile of a top jet differs dramatically from that of a QCD jet \cite{energyprofile}:
the former has a lower $\Psi(r)$ at small $r$ due to the dead-cone effect, but
increases faster with $r$ once the energetic subjets from the weak decay of 
the heavy parent particle start to contribute. A careful look of Fig.~\ref{fig-psi-tot} reveals
a richer structure of the energy profile in the hadronic top
case than in the semi-leptonic top case, which arises form the interplay
between the bottom-quark and $W$-boson contributions with different behaviors
in $r$. This unique feature is attributed to the involved cascade weak hadronic
decays of massive particles.

\begin{center}
 \begin{figure}[htb]
  \def\SCALE{0.27}
  \begin{tabular}{ccc}
   \includegraphics[scale=\SCALE]{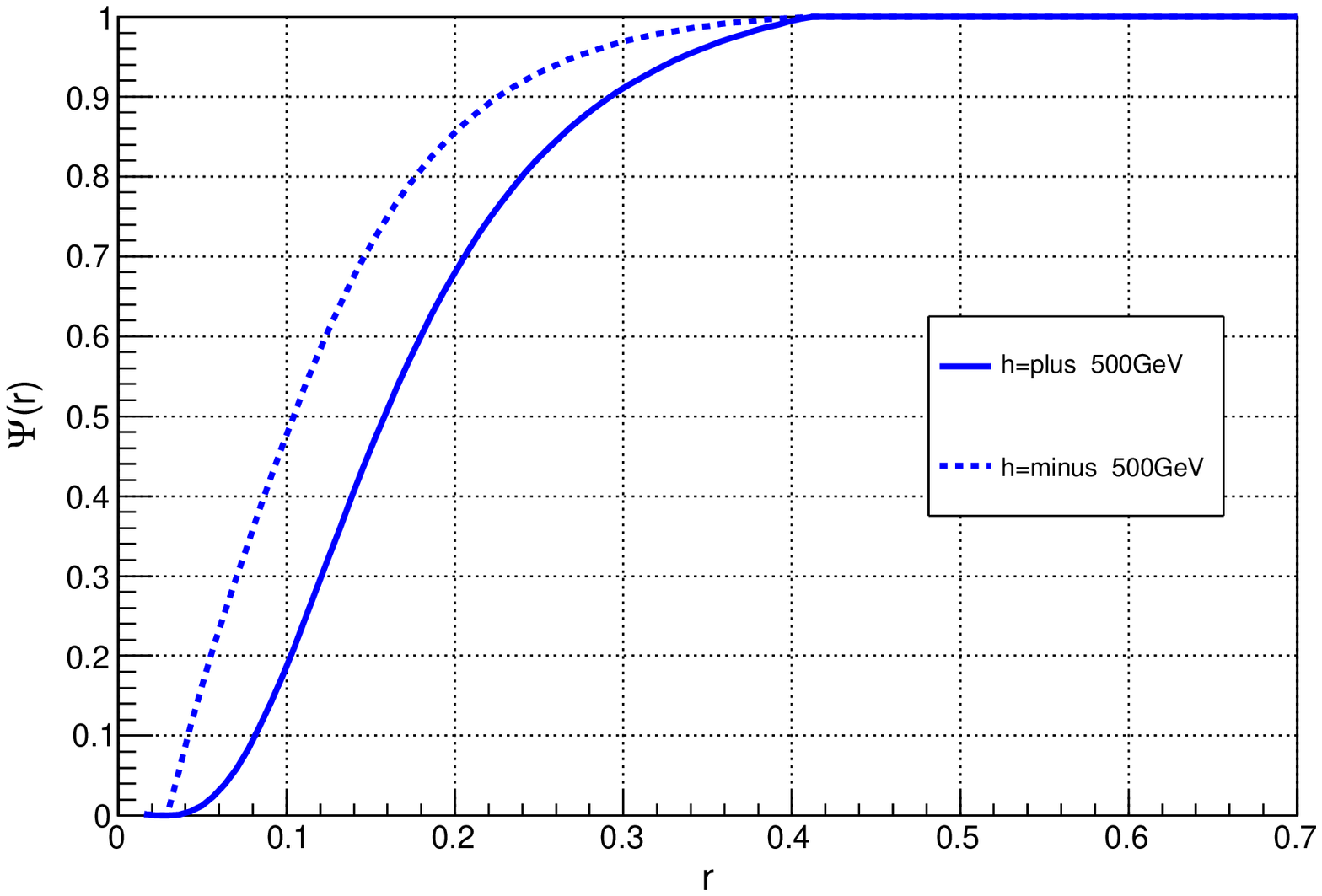} &
   \includegraphics[scale=\SCALE]{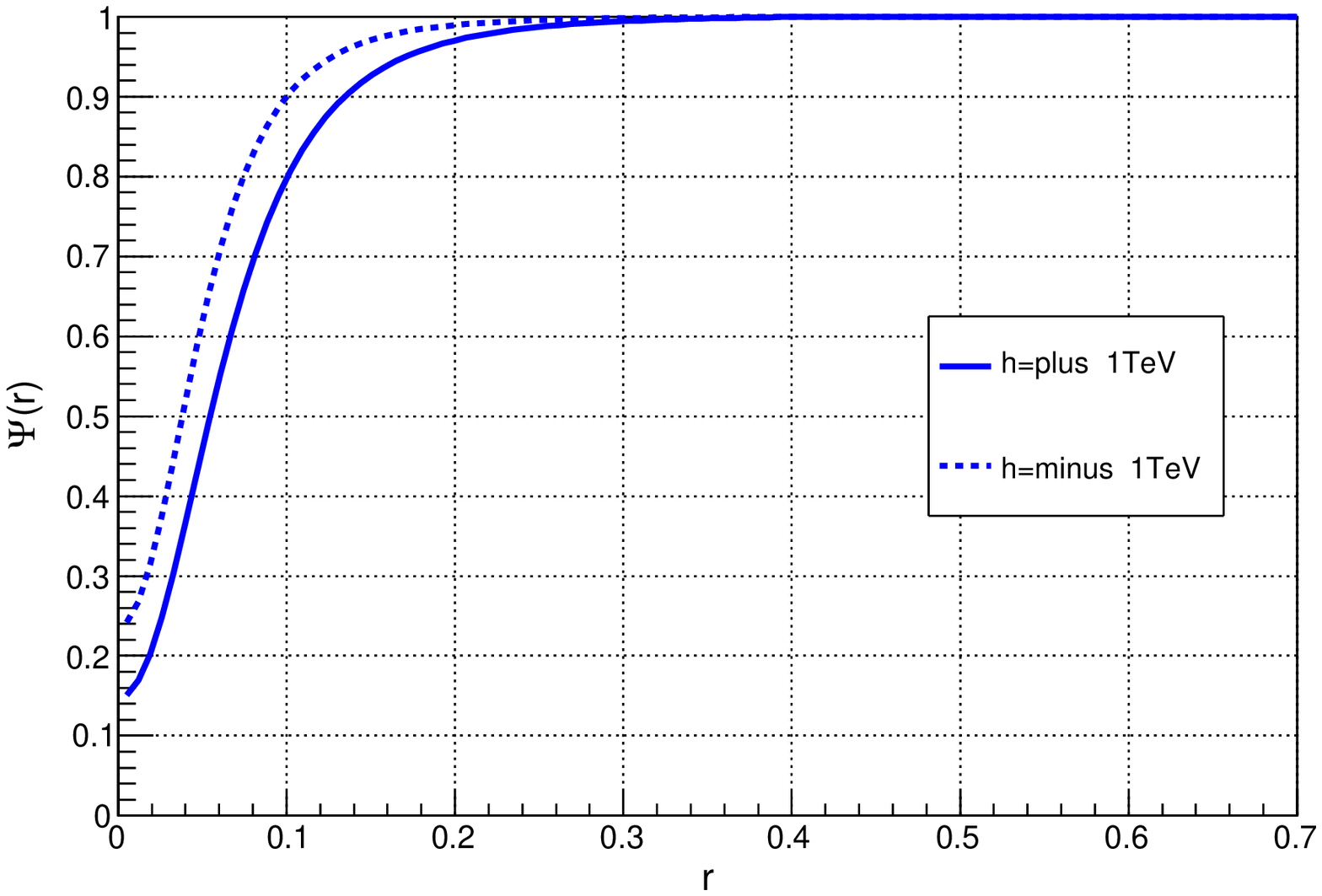} &
   \includegraphics[scale=\SCALE]{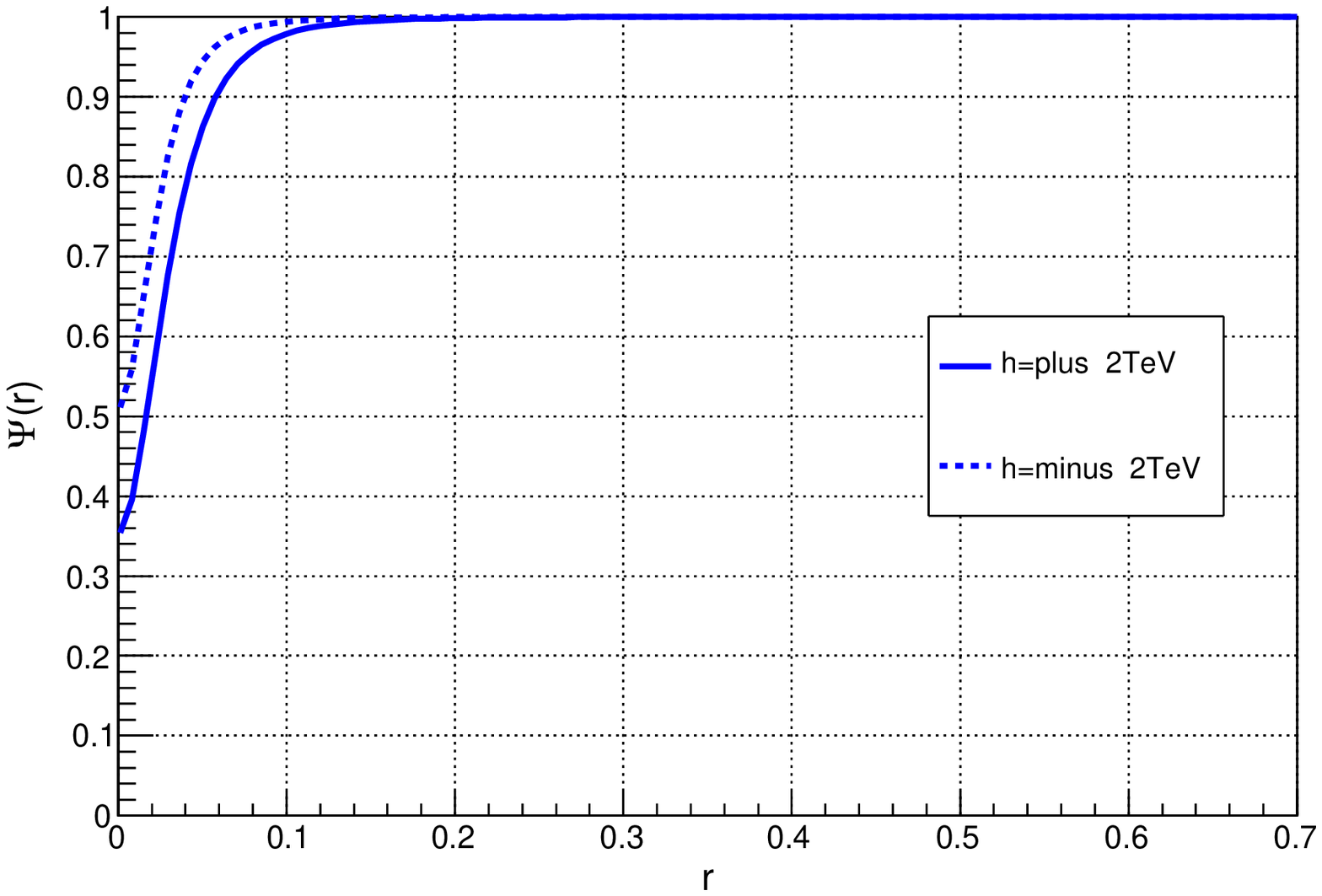} \\
   \hspace{0pt}  (a) &
   \hspace{0pt}  (b) &
   \hspace{0pt}  (c)
  \end{tabular}
  \caption{Bottom jet contributions to the top jet energy profiles for 
  (a) $E_{J_t}=500~\mbox{GeV}$, (b) $E_{J_t}=1~\mbox{TeV}$, and (c) $E_{J_t}=2~\mbox{TeV}$. 
  The top jet radius is set to $R=0.7$.} \label{fig-psi-b}
 \end{figure}
\end{center}

\begin{center}
 \begin{figure}[htb]
  \def\SCALE{0.27}
  \def\OFFSET{20pt}
  \begin{tabular}{ccc}
   \includegraphics[scale=\SCALE]{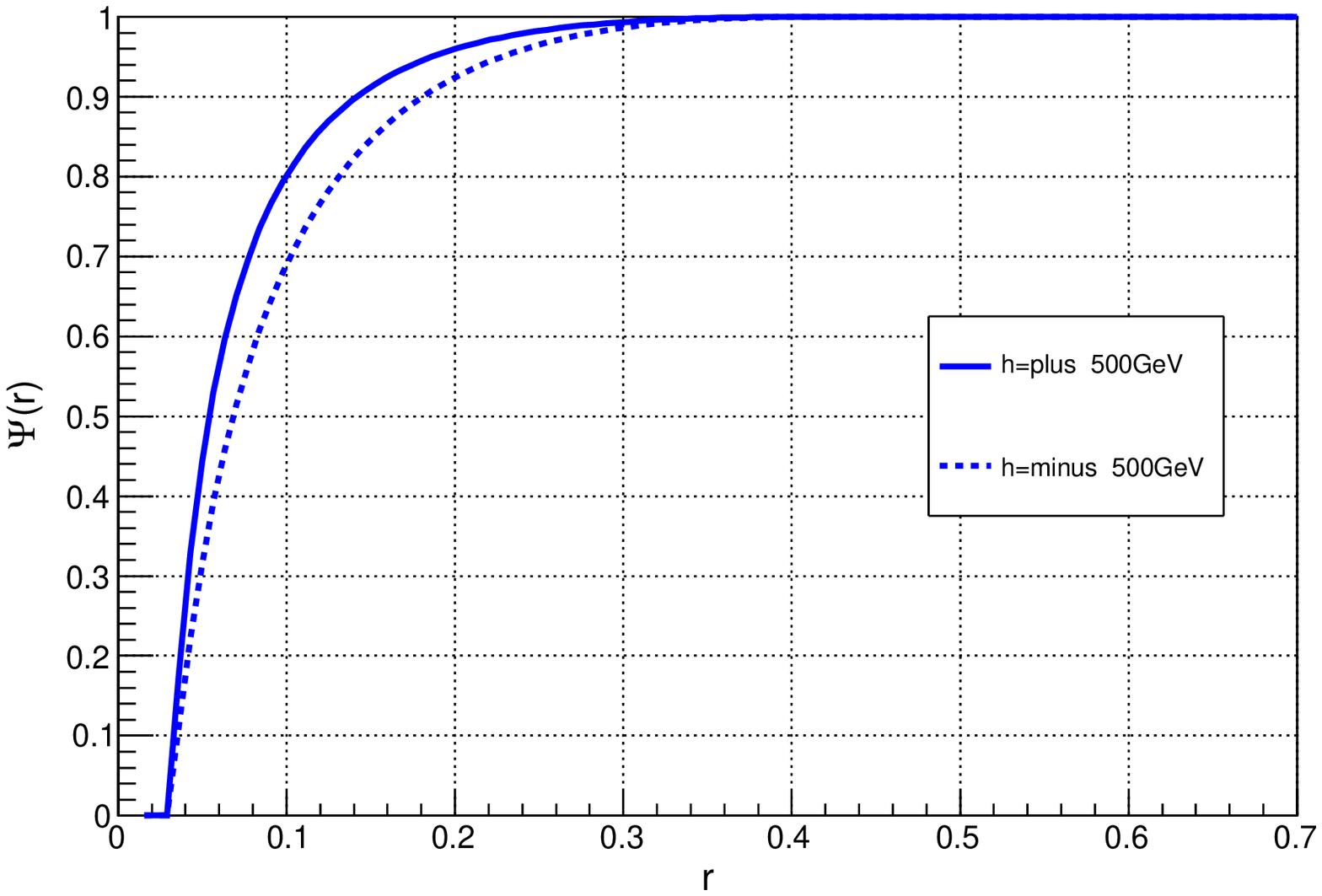} &
   \includegraphics[scale=\SCALE]{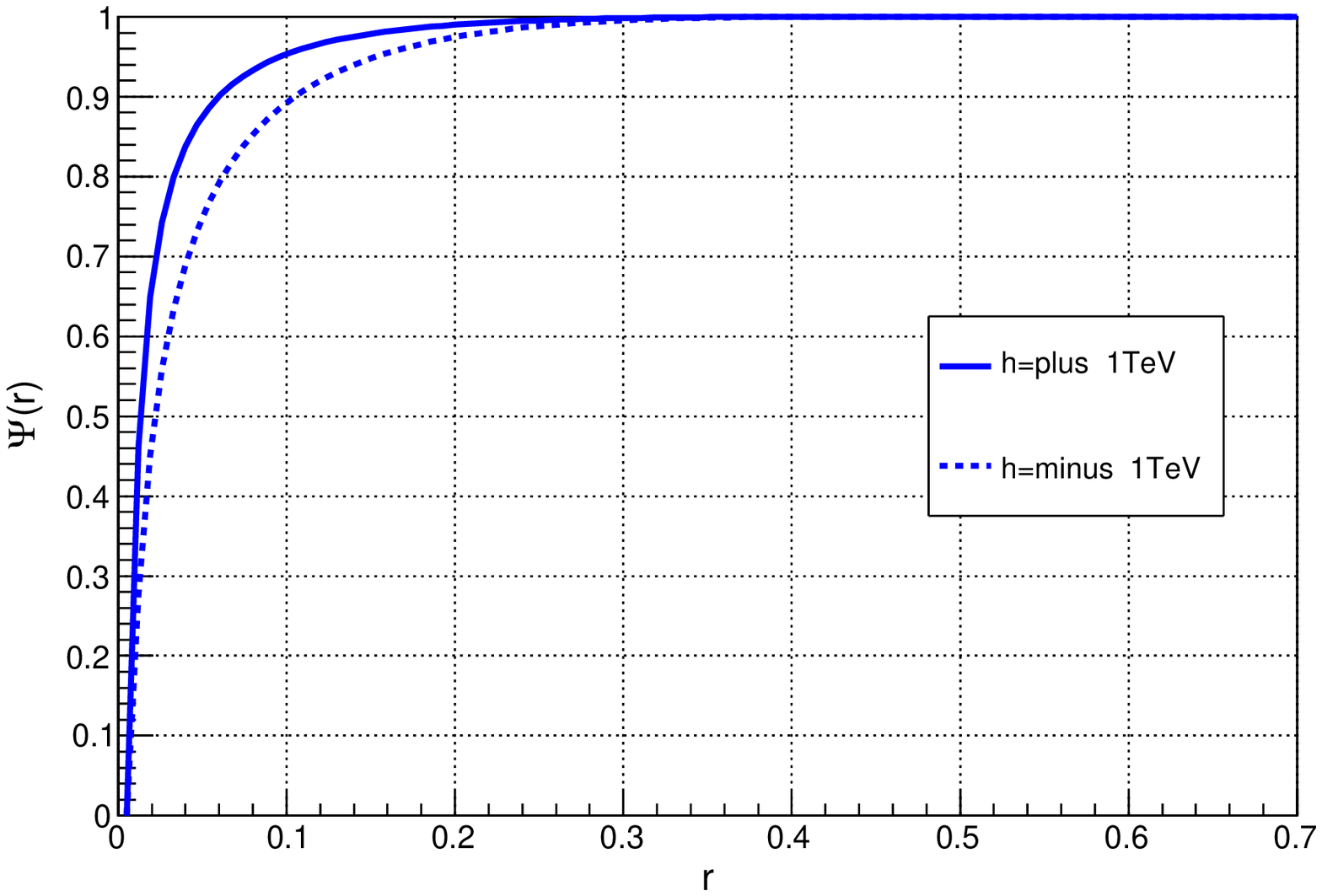} &
   \includegraphics[scale=\SCALE]{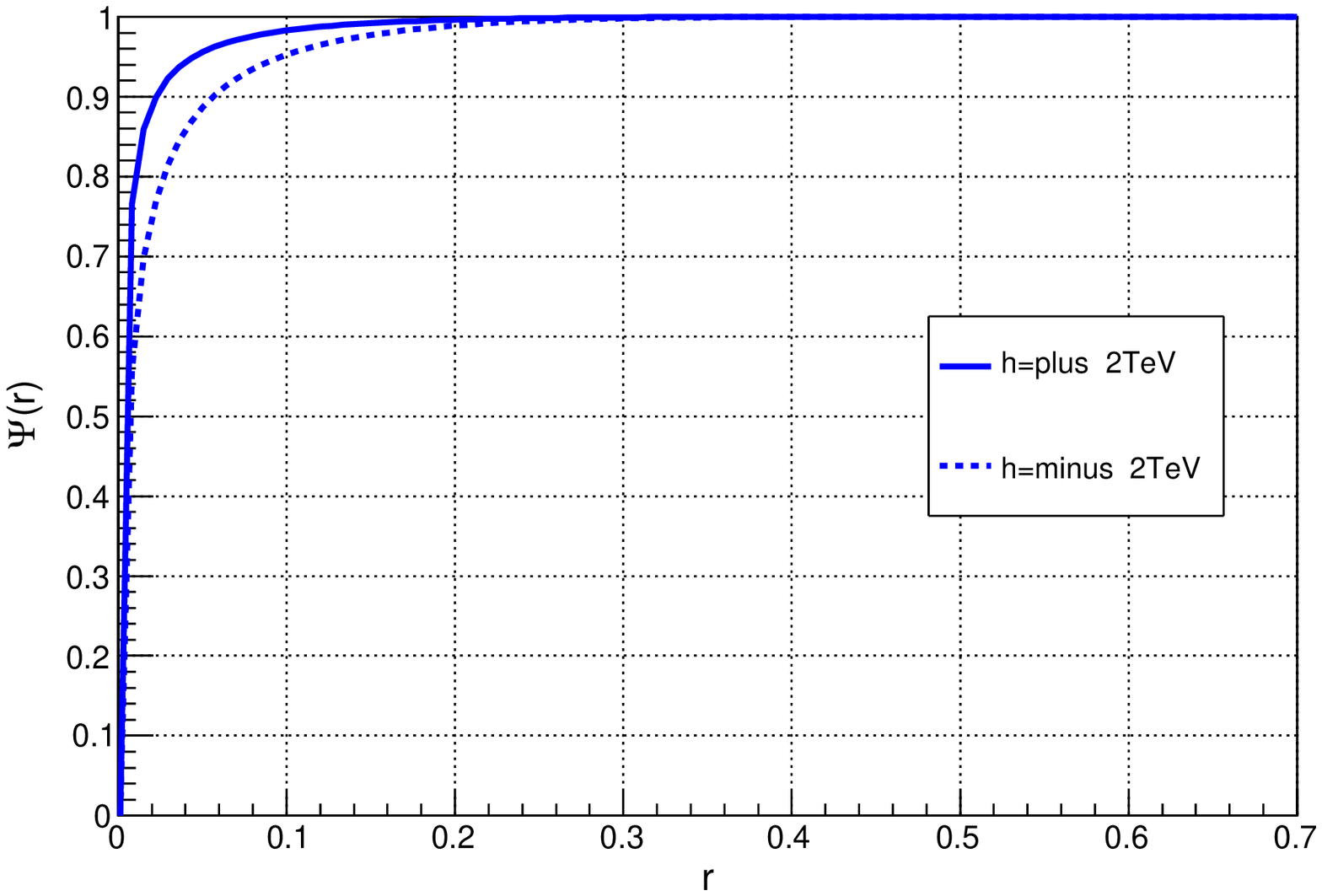} \\
   \hspace{0pt}  (a) &
   \hspace{0pt}  (b) &
   \hspace{0pt}  (c)
  \end{tabular}
  \caption{$W$-boson jet contributions to the top jet energy profiles for 
  (a) $E_{J_t}=500~\mbox{GeV}$, (b) $E_{J_t}=1~\mbox{TeV}$, and (c) $E_{J_t}=2~\mbox{TeV}$. 
  The top jet radius is set to $R=0.7$.} \label{fig-psi-w}
 \end{figure}
\end{center}

\begin{center}
 \begin{figure}[htb]
  \def\SCALE{0.27}
  \def\OFFSET{20pt}
  \begin{tabular}{ccc}
   \includegraphics[scale=\SCALE]{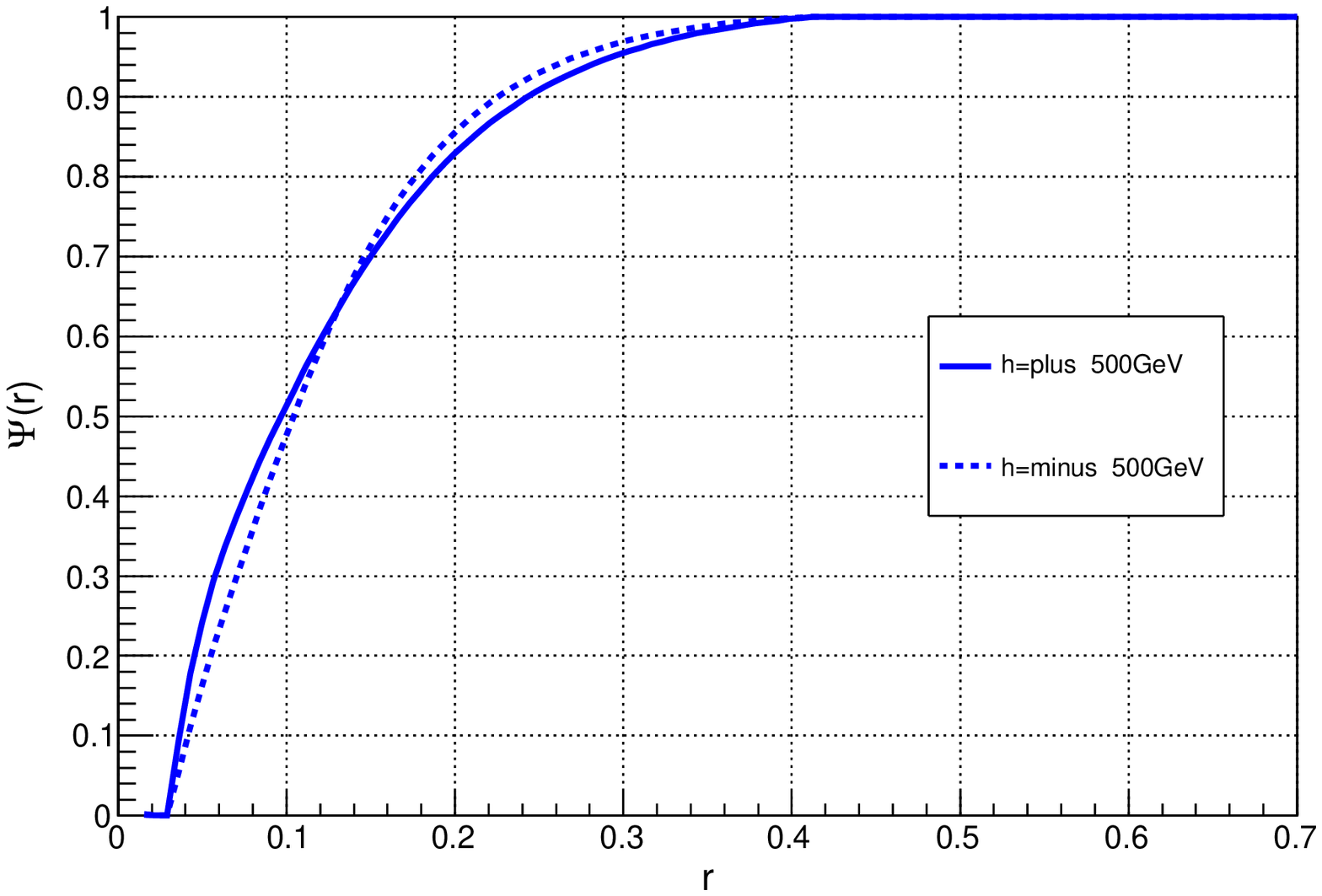} &
   \includegraphics[scale=\SCALE]{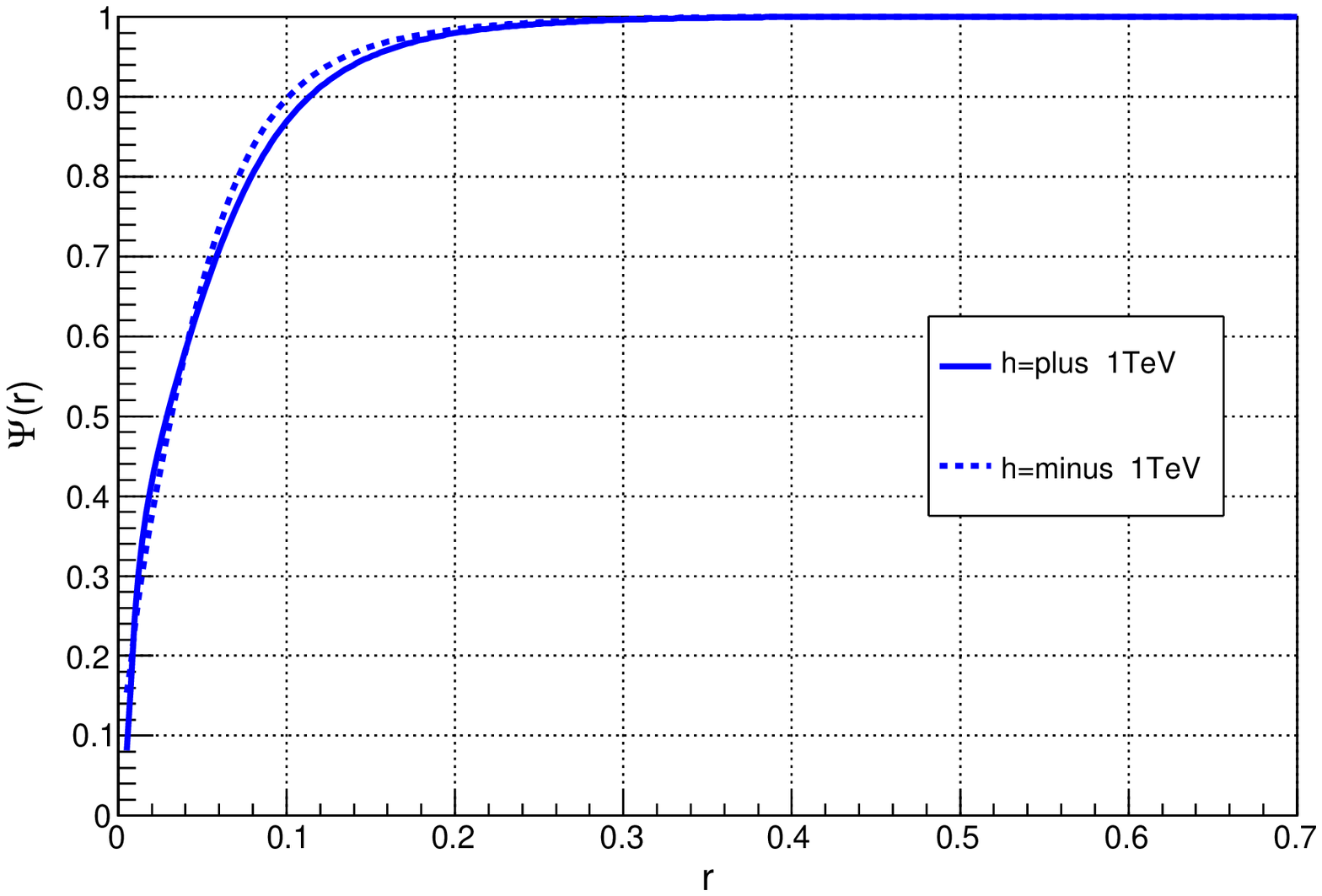} &
   \includegraphics[scale=\SCALE]{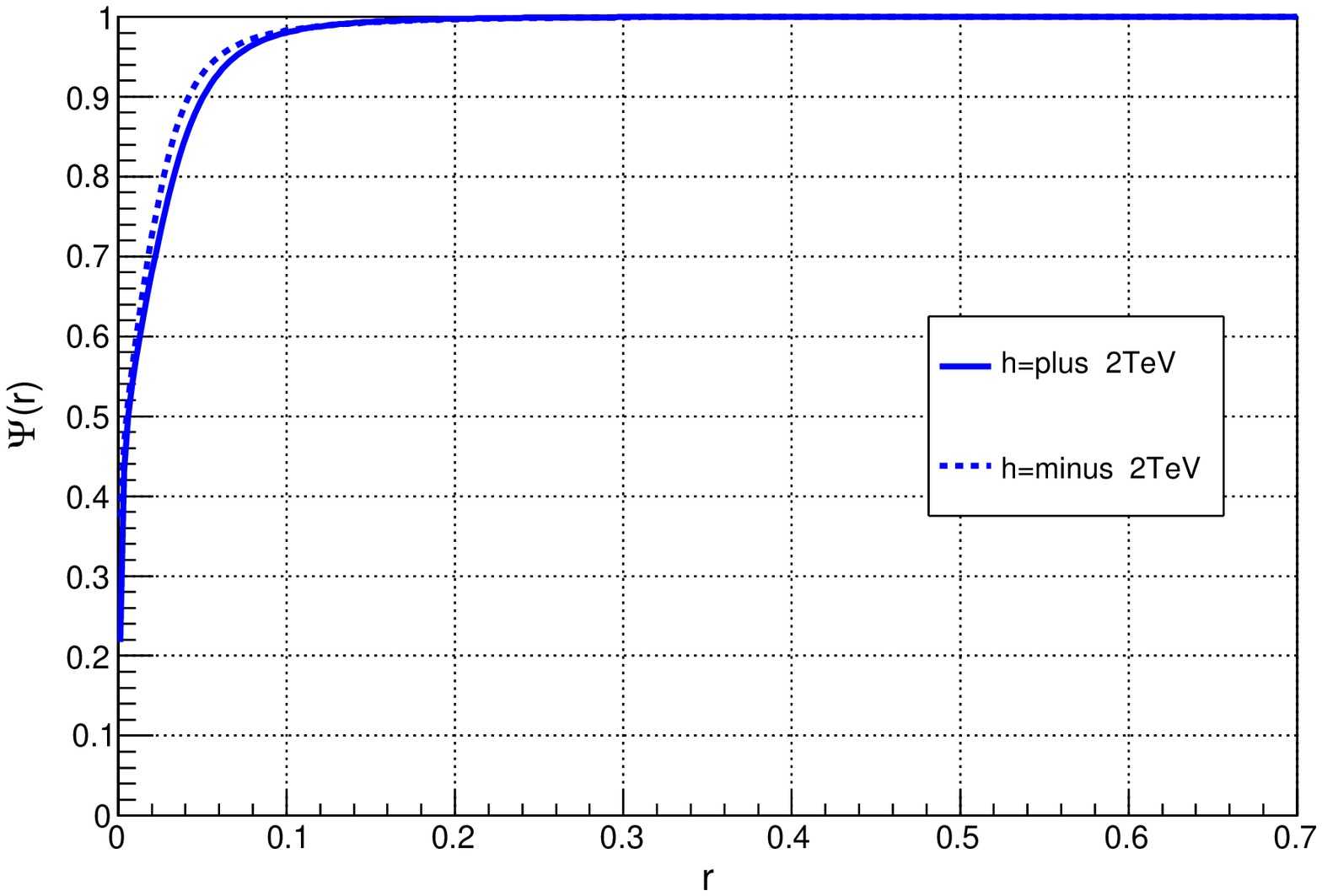} \\
   \hspace{0pt}  (a) &
   \hspace{0pt}  (b) &
   \hspace{0pt}  (c)
  \end{tabular}
  \caption{Energy profiles of hadronic top jets for (a) $E_{J_t}=500~\mbox{GeV}$, 
  (b) $E_{J_t}=1~\mbox{TeV}$, and (c) $E_{J_t}=2~\mbox{TeV}$. The top jet radius 
  is set to $R=0.7$. } \label{fig-psi-tot}
 \end{figure}
\end{center}

To see the slope change in the energy profile more clearly,
we consider the differential jet energy profile $\rho(r)$ defined by
\begin{eqnarray}
 \rho(E_{J_t}, R,r) &\equiv& \frac{d}{dr}\Psi(E_{J_t}, R,r),
\end{eqnarray}
whose results are displayed in Figs.~\ref{fig-diff-psi-tot}(a),
\ref{fig-diff-psi-tot}(b), and \ref{fig-diff-psi-tot}(c) for the top jet energy
$E_{J_t}=500$ GeV, 1 TeV, and 2 TeV, respectively.
The differential jet energy profiles decrease with $r$, similar to what was
observed for QCD jets, but do not follow smooth curves.
The differential jet energy profiles of a right-handed top jet start with larger values,
exhibit more significant drops with $r$, and go below those of a
left-handed top jet. This feature persists even for the top jet energy as high as 2 TeV, 
so the differential jet energy profile can serve as a useful observable for
the helicity identification of a highly boosted hadronic top quark. The curves from the
opposite top helicities become indistinguishable as $r>0.2$ ($r>0.08$, $r>0.03$)
for $E_{J_t}=500$ GeV (1 TeV, 2 TeV).
Since the difference of the jet energy profiles moves toward small $r$, its 
measurement will be challenging when the top jet energy increases. 

\begin{center}
 \begin{figure}[htb]
  \def\SCALE{0.42}
  \def\SCALE{0.27}
  \def\OFFSET{20pt}
  \begin{tabular}{ccc}
   \includegraphics[scale=\SCALE]{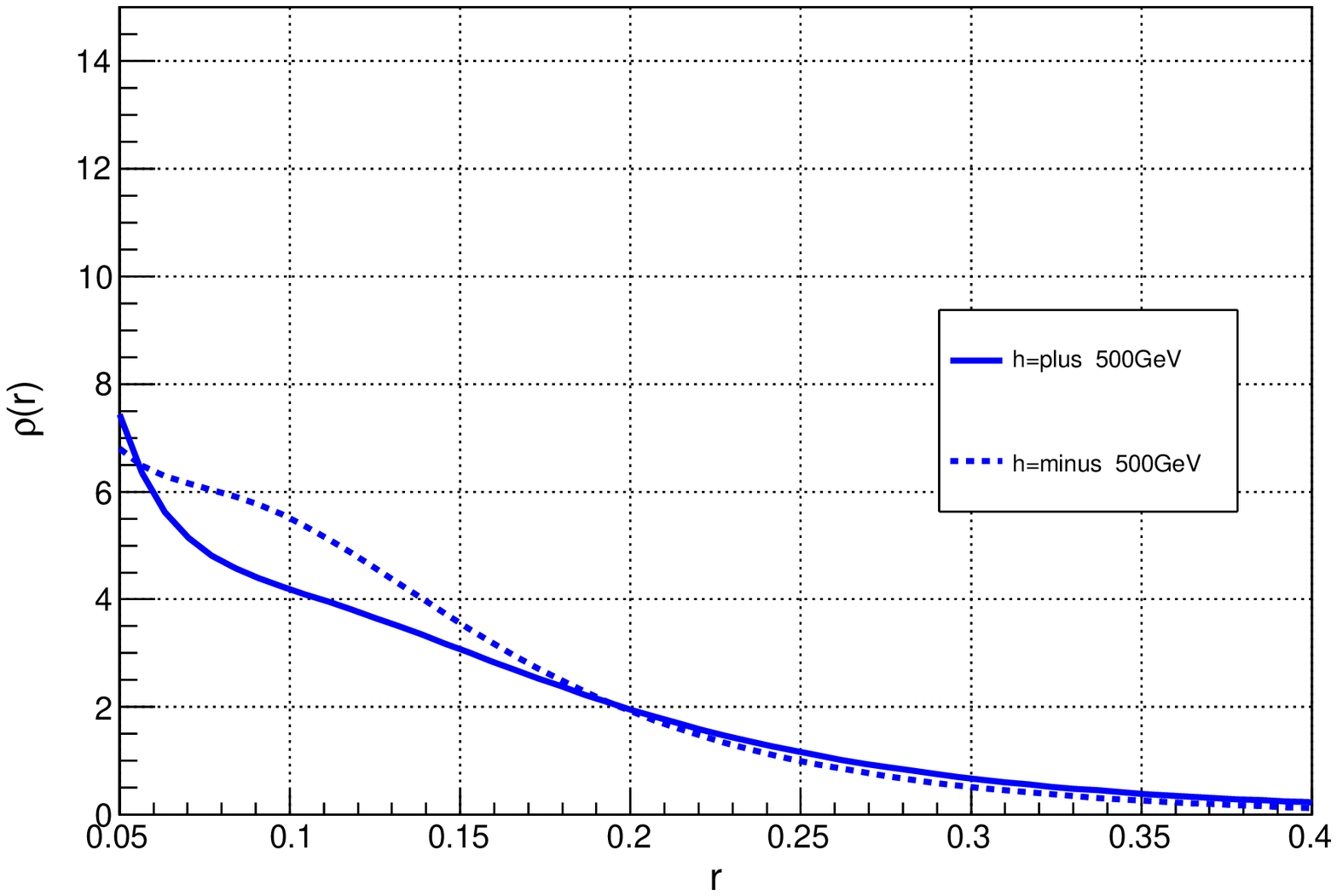} &
   \includegraphics[scale=\SCALE]{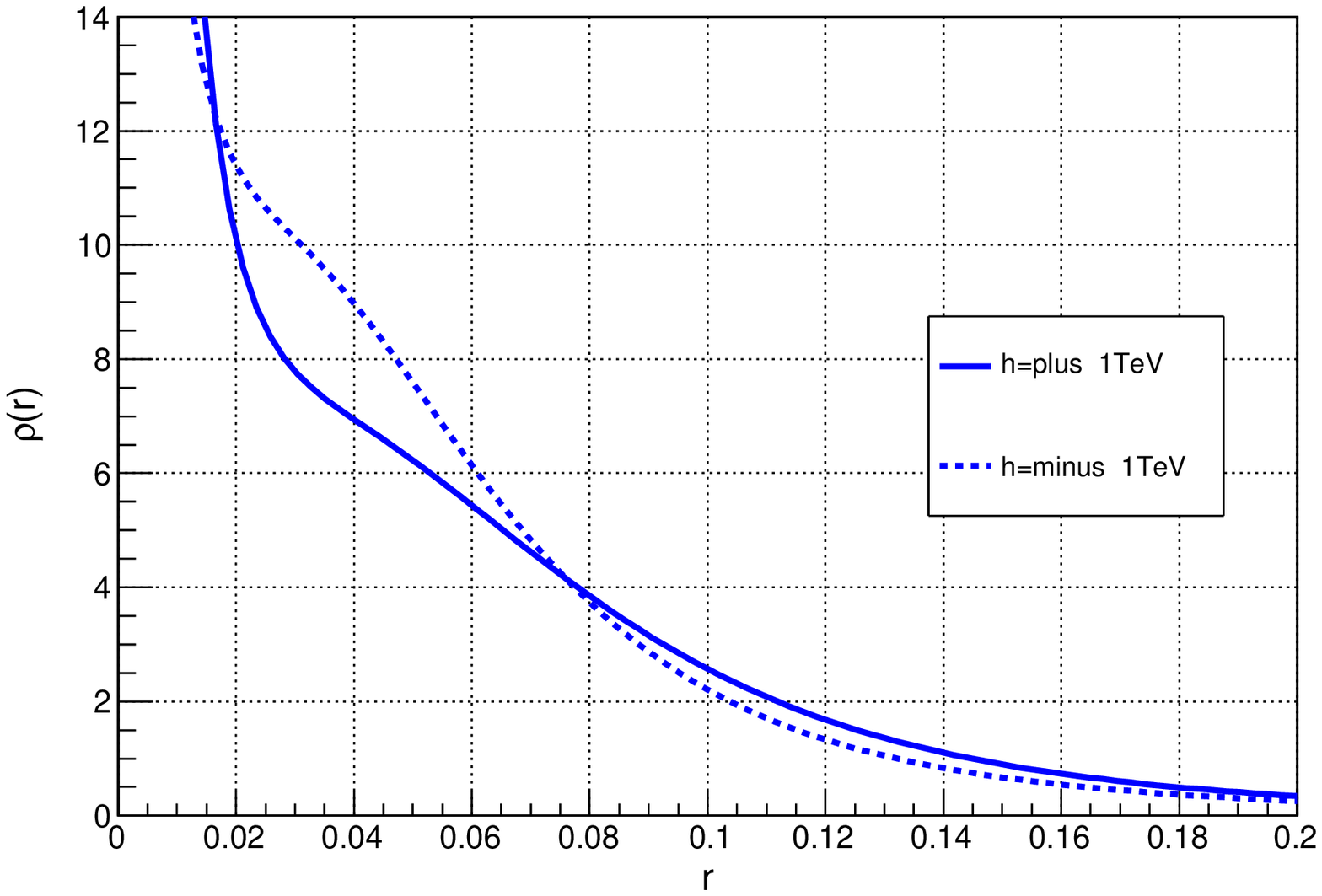} &
   \includegraphics[scale=\SCALE]{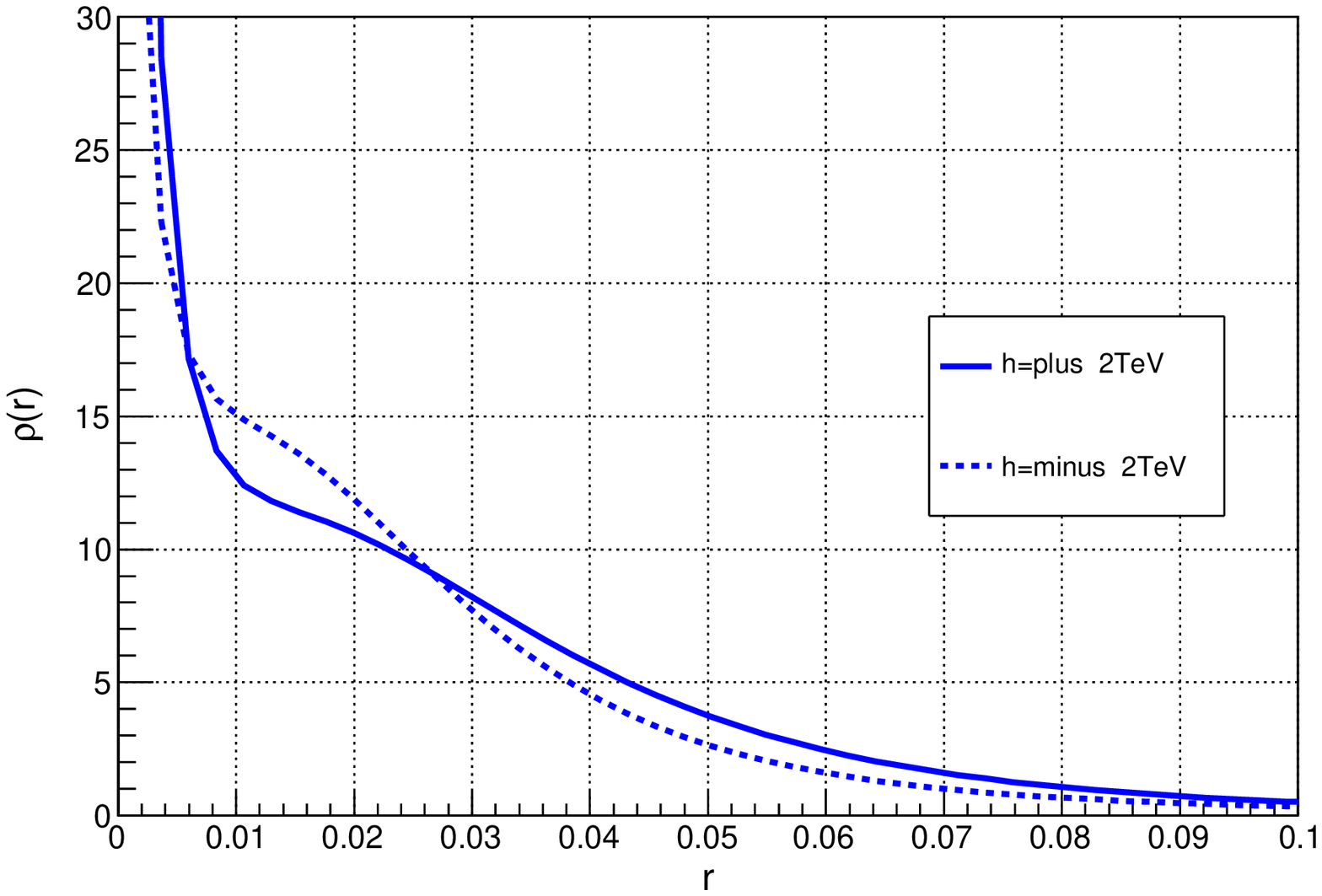}\\
   \hspace{0pt}  (a) &
   \hspace{0pt}  (b) &
   \hspace{0pt}  (c)
  \end{tabular}
  \caption{Differential energy profiles of hadronic top jets for 
  (a) $E_{J_t}=500~\mbox{GeV}$, (b) $E_{J_t}=1~\mbox{TeV}$, and (c) $E_{J_t}=2~\mbox{TeV}$. 
  The top jet radius is set to $R=0.7$.} \label{fig-diff-psi-tot}
 \end{figure}
\end{center}

\section{Conclusion}\label{conclusion}

In this paper we have studied jet substructures of a boosted polarized top quark, which
undergoes the hadronic decay $t\to b u\bar d$, in the pQCD
framework, focusing on the energy profile and the differential
energy profile. These substructures were factorized into the
convolution of a hard top-quark decay kernel with a fat bottom
jet and a fat $W$-boson jet, where the latter is
further factorized into the convolution of a hard $W$-boson decay kernel
with a fat light-quark jet and a thin light-quark jet. In this sequential factorization,
the complicated three-body final-state kinematics was simplified
into two-body one, the soft gluon exchanges among the subjets were absorbed into
the fat subjets, and the jet merging ambiguity was avoided. The pQCD analysis
of the substructures of a hadronic top jet then becomes feasible, and the
formalism presented in this work contributes to the extraction of the
top-quark property from experimental observations.

Computing the heavy-particle decay kernels to LO
in QCD and including the resummation effect in the jet functions,
we have found that the energy profile is not sensitive to the top-quark
helicity because of the compensation between the bottom-quark and $W$-boson
contributions, in which the former (latter) is more dominant in the left-handed
(right-handed) top jet. Instead, the differential energy profile exhibits a richer
structure due to the interplay between the bottom-quark and $W$-boson contributions,
namely, to the involved cascade weak hadronic decays of massive particles,
which also differs dramatically from that of a QCD jet. The differential 
energy profile of a right-handed top jet starts with a larger value,
exhibits a more significant drop with $r$, and becomes lower than that of a
left-handed top jet, a feature which persists
even for the top jet energy as high as 2 TeV. However, this drop is located
at very small $r$, whose observation will be challenging. It is worthwhile
to measure such a feature, which will help to distinguish the helicity of a
highly boosted hadronic top quark, and reveal the chiral structure of physics
beyond the Standard Model.

\begin{acknowledgments}

We thank B. Tweedie and C.P. Yuan for helpful discussions.
This work was supported in part
by the Ministry of Science and Technology of R.O.C. under Grant No.
NSC-104-2112-M-001-037-MY3.

\end{acknowledgments}

\appendix
\section{SOFT-GLUON CONTRIBUTION AT ONE LOOP}

In this appendix we calculate soft-gluon corrections to a hadronic top jet
at one loop, and demonstrate that their contributions can be absorbed into
the fat subjets in the factorization of the hadronic top jet. First, a soft
gluon emitted by or attaching to the hard top-quark decay kernel leads to
power-suppressed correction, which will be neglected here. Soft gluons
emitted by other lines (the light quarks and the Wilson links associated with
the top jet definition) can be factorized through the standard eikonal
approximation into a nonperturbative soft function $S(\omega)$, $\omega$ being
proportional to the soft gluon energy \cite{boost.higgs.subjet}. The soft function from
virtual and real gluons exchanged between two pairs of eikonal lines in the directions
$\bar\xi_1$ and $\bar\xi_2$ within the top jet cone is given, at one loop, by
\cite{boost.higgs.subjet}
\begin{eqnarray}
S^{(1)}=\frac{\alpha_s C_F}{\pi(R E_{J_t})^2}\ln\frac
{\bar\xi_{1}^2\bar\xi_{2}^2}
{4(\bar\xi_{1}\cdot \bar\xi_{2})^2}\left(\frac{1}{\epsilon}
+\ln\frac{4\pi\mu_f^2\bar N^2}{R^4E_{J_t}^2e^{\gamma_E}}\right),
\label{stt}
\end{eqnarray}
where $\mu_f$ is a factorization scale, the color factor $C_F=4/3$ and the moment $\bar N\equiv
N\exp(\gamma_E)$, with $\gamma_E$ being the Euler constant.
As $\bar\xi_{i}$ represents the direction of light subjet $i$, the invariant $\bar\xi_{i}^2$
implies that $S^{(1)}$ contains the collinear dynamics which
has been absorbed into the light subjet function $J_i$. Collinear subtraction
from the soft function is thus necessary for avoiding double counting of collinear
dynamics.

In the semi-leptonic top jet soft gluons are exchanged only between the bottom quark
and the Wilson line along the direction $\xi_t$ that appears in the top-jet
definition at leading power. Therefore, the soft correction $S_{b\xi}^{(1)}$ is obtained 
by the substitutions $\bar\xi_{1}\to \bar\xi_{J_b}$ and $\bar\xi_{2}\to \xi_{J_t}$ in Eq.~(\ref{stt}). 
The bottom jet in the semi-leptonic top jet is chosen as a fat jet of cone radius $R$
\cite{Kitadono:2014hna}. The collinear subtraction
term to be deducted from Eq.~(\ref{stt}) is written as
\begin{eqnarray}
S_b^{(1)}=\frac{\alpha_s C_F}{\pi(R E_{J_t})^2}\ln\frac
{\bar\xi_{J_b}^2\xi_{J_b}^2}
{4(\bar\xi_{J_b}\cdot \xi_{J_b})^2}\left(\frac{1}{\epsilon}
+\ln\frac{4\pi\mu_f^2\bar N^2}{R^4E_{J_t}^2e^{\gamma_E}}\right),
\label{st1}
\end{eqnarray}
where $\xi_{J_b}$ denotes the direction of the Wilson line in the bottom jet definition.
Because $\xi_{J_t}$ may not be close to the light cone in the top jet construction, 
no collinear subtraction associated with this vector is needed. 

The combination of Eqs.~(\ref{stt}) and (\ref{st1}) yields
\begin{eqnarray}
S_{b\xi}^{(1)}-S_b^{(1)}=\frac{\alpha_s C_F}{\pi(R E_{J_t})^2}\ln\frac
{R^2\xi_{J_t}^2}
{4(\bar\xi_{J_b}\cdot \xi_{J_t})^2}\left(\frac{1}{\epsilon}
+\ln\frac{4\pi\mu_f^2\bar N^2}{R^4E_{J_t}^2e^{\gamma_E}}\right),
\label{st2}
\end{eqnarray}
in which we have further imposed the condition
$4(\bar\xi_{J_b}\cdot \xi_{J_b})^2/\xi_{J_b}^2=R^2$ for defining a quark jet
in the resummation formalism \cite{energyprofile}. With the similar condition for the
top jet, $4(\bar\xi_{J_t}\cdot \xi_{J_t})^2/\xi_{J_t}^2=R^2$, it is trivial to show that
the logarithmic coefficient $\ln\{R^2\xi_{J_t}^2/[4(\bar\xi_{J_b}\cdot \xi_{J_t})^2]\}$
is proportional to the polar angle $\theta_{J_b}$ of the bottom jet.
Equation~(\ref{thetab}) then implies that Eq.~(\ref{st2}) represents a power-suppressed
correction, and is negligible. That is, soft gluons have been absorbed into the fat
bottom jet in the semi-leptonic top jet, and the remaining soft corrections are
of higher powers. In the hadronic top jet, the soft gluons exchanged between the bottom
quark and the Wilson line in the direction $\xi_{J_t}$ within the top cone of radius $R$ is
the same as $S_{b\xi}^{(1)}$. The bottom jet is chosen as a fat one with the cone
radius $R$ in this case, so the collinear subtraction term is also the same as in
Eq.~(\ref{st2}). The subtracted soft correction $S_{b\xi}^{(1)}-S_b^{(1)}$ is thus
negligible.

It has been demonstrated that the soft gluons exchanged between the fat subjet
and the thin subjet in the $W$-boson jet can be absorbed into the fat subjet
\cite{boost.higgs.subjet}. Here we quoted the one-loop soft correction
and the collinear subtraction terms with the up jet being a fat subjet,
and the down jet being a thin subjet,
\begin{eqnarray}
S_{ud}^{(1)}&=&\frac{\alpha_s C_F}{\pi(R E_{J_t})^2}\ln\frac
{\bar\xi_{J_u}^2\bar\xi_{J_d}^2}
{4(\bar\xi_{J_u}\cdot \bar\xi_{J_d})^2}\left(\frac{1}{\epsilon}
+\ln\frac{4\pi\mu_f^2\bar N^2}{R^4E_{J_t}^2e^{\gamma_E}}\right),\nonumber\\
S_u^{(1)}&=&\frac{\alpha_s C_F}{\pi(R E_{J_t})^2}\ln\frac
{\bar\xi_{J_u}^2}{R^2}\left(\frac{1}{\epsilon}
+\ln\frac{4\pi\mu_f^2\bar N^2}{R^4E_{J_t}^2e^{\gamma_E}}\right),\nonumber\\
S_d^{(1)}&=&\frac{\alpha_s C_F}{\pi(R E_{J_t})^2}\ln\frac
{R^2\bar\xi_{J_d}^2}
{4(\bar\xi_{J_u}\cdot \bar\xi_{J_d})^2}\left(\frac{1}{\epsilon}
+\ln\frac{4\pi\mu_f^2\bar N^2}{R^4E_{J_t}^2e^{\gamma_E}}\right).
\label{wtt}
\end{eqnarray}
The thin down jet with the cone radius $r$ has a low invariant mass, whose
dependence can be dropped in all other subprocesses of the factorization formula.
This mass can then be integrated out trivially, and a thin subjet contributes an
overall normalization to the factorization formula. Hence, the condition
$4(\bar\xi_{J_d}\cdot \xi_{J_d})^2/\xi_{J_d}^2=r^2$ required by the resummation formalism
\cite{energyprofile} is relaxed, and the Wilson line direction
$\xi_{J_d}$ can be arbitrary in principle. We utilize this freedom, choosing
$\xi_{J_d}$ to give the logarithmic coefficient of $S_d^{(1)}$ in Eq.~(\ref{wtt}).
This choice is feasible, because of $(\bar\xi_{J_u}\cdot \bar\xi_{J_d})^2\sim (m_W/E_{J_t})^4
\sim O(r^2)$ in the considered kinematic region.
The subtracted soft correction $S_{ud}^{(1)}-S_u^{(1)}-S_d^{(1)}$ is then
negligible.

Next we discuss the soft gluons exchanged between the $W$-boson jet and the
Wilson line of the top jet. Those between the fat subjet and the
Wilson line of the top jet, and those between the thin subjet and the Wilson
line yield
\begin{eqnarray}
S_{u\xi}^{(1)}&=&\frac{\alpha_s C}{\pi(R E_{J_t})^2}\ln\frac
{\bar\xi_{J_u}^2\xi_{J_t}^2}
{4(\bar\xi_{J_u}\cdot \xi_{J_t})^2}\left(\frac{1}{\epsilon}
+\ln\frac{4\pi\mu_f^2\bar N^2}{R^4E_{J_t}^2e^{\gamma_E}}\right),\nonumber\\
S_{d\xi}^{(1)}&=&-\frac{\alpha_s C}{\pi(R E_{J_t})^2}\ln\frac
{\bar\xi_{J_d}^2\xi_{J_t}^2}
{4(\bar\xi_{J_d}\cdot \xi_{J_t})^2}\left(\frac{1}{\epsilon}
+\ln\frac{4\pi\mu_f^2\bar N^2}{R^4E_{J_t}^2e^{\gamma_E}}\right),
\end{eqnarray}
respectively. The coefficient $C$ differs from $C_F$, because a light quark
in the $W$-boson jet and the Wilson line of the top jet are in separated color flows.
Its explicit value is not crucial here. The minus sign in $S_{d\xi}^{(1)}$ arises,
since the down quark is an anti-quark. It is straightforward to show, for the
same collinear regulators $\bar\xi_{J_u}^2=\bar\xi_{J_d}^2$, that
the combination $S_{u\xi}^{(1)}+S_{d\xi}^{(1)}$ is proportional to the polar angles
$\theta_{J_u}$ and $\theta_{J_d}$, which are power-suppressed in an energetic top jet
as indicated in Eq.~(\ref{thetad}) and by the perpendicular momentum conservation
$E_{J_u}\theta_{J_u}=E_{J_{\bar d}}\theta_{J_{\bar d}}$. This observation realizes
the postulation in the Introduction that soft gluon exchanges between
the color-singlet $W$-boson jet and other subprocesses are expected to be suppressed
in the limit of high jet energy. This soft correction does not double count
the leading-power collinear dynamics, so the collinear subtraction is not
necessary.

At last, the soft gluons exchanged among the up, down, and bottom subjets are
handled in the similar way. We have the one-loop soft corrections
\begin{eqnarray}
S_{ub}^{(1)}&=&\frac{\alpha_s C}{\pi(R E_{J_t})^2}\ln\frac
{\bar\xi_{J_u}^2\bar\xi_{J_b}^2}
{4(\bar\xi_{J_u}\cdot \bar\xi_{J_b})^2}\left(\frac{1}{\epsilon}
+\ln\frac{4\pi\mu_f^2\bar N^2}{R^4E_{J_t}^2e^{\gamma_E}}\right),\nonumber\\
S_{db}^{(1)}&=&-\frac{\alpha_s C}{\pi(R E_{J_t})^2}\ln\frac
{\bar\xi_{J_d}^2\bar\xi_{J_b}^2}
{4(\bar\xi_{J_d}\cdot \bar\xi_{J_b})^2}\left(\frac{1}{\epsilon}
+\ln\frac{4\pi\mu_f^2\bar N^2}{R^4E_{J_t}^2e^{\gamma_E}}\right),
\end{eqnarray}
whose combination $S_{ub}^{(1)}+S_{db}^{(1)}$ is also power-suppressed.
We conclude that the soft gluons are factorized into the fat
subjets in our construction at least at one-loop level, and expect that such a
factorization scheme can be extended to all orders.


\end{document}